\documentclass[a4paper,usenatbib]{mnras}

\usepackage{newtxtext,newtxmath}

\usepackage[T1]{fontenc}
\usepackage{ae,aecompl}

\usepackage{url}
\usepackage{lscape}
\usepackage{longtable}
\usepackage{footnote}
\usepackage{tablefootnote}
\usepackage{booktabs}
\usepackage{hyperref}
\usepackage[hyphenbreaks]{breakurl}
\usepackage{cleveref}

\usepackage{graphicx}   
\usepackage{amsmath, amsfonts}  

\newcommand{\Lagr}{\mathcal{L}}

\def\la{\mathrel{\rlap{\lower4pt\hbox{\hskip1pt$\sim$}}
    \raise1pt\hbox{$<$}}}                
\def\ga{\mathrel{\rlap{\lower4pt\hbox{\hskip1pt$\sim$}}
    \raise1pt\hbox{$>$}}}                

\def\plotonekindaspecial#1{\centering \leavevmode
    \includegraphics[angle=0,width=2.2\columnwidth]{#1}}

\def\plottwo#1#2{\centering \leavevmode
    \includegraphics[angle=0,width=0.98\columnwidth]{#1} \hfil
    \includegraphics[angle=0,width=0.98\columnwidth]{#2}}

\def\plottwoalmostspecial#1#2{\centering \leavevmode
    \includegraphics[angle=0,width=1.00\columnwidth]{#1} \hfil
    \includegraphics[angle=0,width=0.97\columnwidth]{#2}}

\def\plottwospecial#1#2{\centering \leavevmode
    \includegraphics[angle=0,width=0.96\columnwidth]{#1} \hfil
    \includegraphics[angle=0,width=0.99\columnwidth]{#2}}

\def\plotthree#1#2#3{\centering \leavevmode
    \includegraphics[angle=0,width=0.97\columnwidth]{#1} \hfil
    \includegraphics[angle=0,width=0.97\columnwidth]{#2} \hfil
    \includegraphics[angle=0,width=0.97\columnwidth]{#3} }

\def\plotone#1{\centering \leavevmode
    \includegraphics[angle=0,width=1.0\columnwidth]{#1}}

\def\orcid#1#2{\href{https://orcid.org/#2}{{#1} \includegraphics{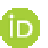}}}

\title[Ly$\alpha$-UV Offsets in Galaxies at $z\sim6$]{The Size and Pervasiveness of Ly$\alpha$-UV Spatial Offsets in Star-Forming Galaxies at $z\sim6$}
\author[B.~C.\ Lemaux et al.]{\orcid{B.~C.\ Lemaux}{0000-0002-1428-7036}$^{1}$\thanks{E-mail: blemaux@ucdavis.edu}, \orcid{S.\ Fuller}{0000-0001-9884-4807}$^{1}$, 
\orcid{M.\ Brada\v{c}}{0000-0001-5984-0395}$^{1}$, \orcid{L.\ Pentericci}{0000-0001-8940-6768}$^{2}$, \orcid{A.\ Hoag}{0000-0001-8989-2567}$^{3}$, \newauthor 
\orcid{V.\ Strait}{0000-0002-6338-7295}$^{1}$,\orcid{T.\ Treu}{0000-0002-8460-0390}$^3$, C.\ Alvarez$^{4}$, \orcid{P.\ Bolan$^{1}$}{0000-0002-7365-4131}, 
\orcid{P.~J.\ Gandhi$^{1}$}{0000-0003-0965-605X}, 
\orcid{K.-H.\ Huang}{0000-0001-7826-6448}$^{1}$, \newauthor 
\orcid{T. Jones}{0000-0001-5860-3419}$^1$, \orcid{C.\ Mason}{0000-0002-3407-1785}$^{5,6}$, \orcid{D.\ Pelliccia}{0000-0002-3007-0013}$^{1,7}$, 
\orcid{B.\ Ribeiro}{0000-0002-4844-0414}$^{8}$, \orcid{R.~E.\ Ryan}{0000-0003-0894-1588}$^{9}$, \newauthor 
\orcid{K.~B.\ Schmidt}{0000-0002-3418-7251}$^{10}$, \orcid{E.\ Vanzella}{0000-0002-5057-135X}$^{11}$, \orcid{Y.\ Khusanova}{0000-0002-7220-397X}$^{12,13}$, 
\orcid{O.\ Le F\`{e}vre}{0000-0001-5891-2596}$^{13}$, \newauthor \orcid{L.\ Guaita}{0000-0002-4902-0075}$^{2,14}$, \orcid{N.~P.\ Hathi}{0000-0001-6145-5090}$^{9}$, 
\orcid{A.\ Koekemoer}{0000-0002-6610-2048}$^{9}$, \orcid{J.\ Pforr}{0000-0002-3414-8391}$^{15}$  
$^{\emph{\small{\rm{(Affiliations are listed at the end of the paper)}}}}$} 

\date{Received: June 29th, 2020; Revised: March 1st, 2021; Accepted: March 24th, 2021}

\pubyear{2021}

\begin{document}
\label{firstpage}
\pagerange{\pageref{firstpage}--\pageref{lastpage}}
\maketitle
%

\begin{abstract}
We study the projected spatial offset between the ultraviolet continuum and Ly$\alpha$ emission for 65 lensed and 
unlensed galaxies in the Epoch of Reionization ($5\le z \le 7$), the first such study at these redshifts, in order to understand the potential 
for these offsets to confuse estimates of the Ly$\alpha$ properties of galaxies observed in slit spectroscopy. While we find that $\sim$40\% of galaxies in our sample
show significant projected spatial offsets ($|\Delta_{\rm{Ly}\alpha-\rm{UV}}|$), we find a relatively modest average projected offset of 
$|\widetilde{\Delta}_{\rm{Ly}\alpha-\rm{UV}}|$ = 0.61$\pm$0.08 proper kpc for the entire sample. A small fraction of our sample, $\sim$10\%, exhibit offsets
in excess of 2 proper kpc, with offsets seen up to $\sim$4 proper kpc, sizes that are considerably larger than the effective radii of typical galaxies at 
these redshifts. An internal comparison and a comparison to 
studies at lower redshift yielded no significant evidence of evolution of $|\Delta_{\rm{Ly}\alpha-\rm{UV}}|$ with redshift. In our sample, 
UV-bright galaxies ($\widetilde{L_{UV}}/L^{\ast}_{UV}=0.67$) showed offsets a factor of three greater than their fainter counterparts 
($\widetilde{L_{UV}}/L^{\ast}_{UV}=0.10$), 0.89$\pm$0.18 vs.\ 0.27$\pm$0.05 proper kpc, respectively. The presence of companion galaxies and early-stage 
merging activity appeared to be unlikely causes of these offsets.
Rather, these offsets appear consistent with a scenario in which internal anisotropic processes resulting from stellar feedback, 
which is stronger in UV-brighter galaxies, facilitate 
Ly$\alpha$ fluorescence and/or backscattering from nearby or outflowing gas. The reduction in the Ly$\alpha$ flux due to offsets 
was quantified. It was found that the differential loss of Ly$\alpha$ photons for 
galaxies with average offsets is not, if corrected for, a limiting factor for all but the narrowest slit widths ($<$0.4$\arcsec$).
However, for the largest offsets, if they are mostly perpendicular to the slit major axis, slit losses were found to be extremely severe in 
cases where slit widths of $\le$1$\arcsec$ were employed, such as those planned for \emph{James Webb Space Telescope}/NIRSpec observations. 

\end{abstract}




\section{Introduction}

The advent of ground-based 8-10-\emph{m} class telescopes such as the Very Large Telescope array (VLT), Keck, Gemini, and Subaru, extremely sensitive ground-based sub-millimeter arrays such 
as the Atacama Large Millimeter/submillimeter Array (ALMA), as well as space-based optical/near-infrared facilities such as the \emph{Hubble} 
Space Telescope (\emph{HST}) and the \emph{Spitzer} Space Telescope, performing imaging and spectroscopy at a variety of wavelengths have enabled studies of galaxies at increasingly 
higher redshifts. These studies, especially those over the last decade, have now become powerful enough to begin to explore the epoch known as 
``reionization'' (EoR), during which the first stars, galaxies, and active galactic nuclei (AGN) were beginning to ionize what was a neutral intergalactic medium of hydrogen gas 
(e.g., \citealt{treu13, tilvi14, tilvi20, schmidt16, hashimoto18, hashimoto18b, hashimoto19, mason18, mason19, austin19a, tamura19, spencer20, pellixDP7}). It is expected that the very first 
galaxies started to form when the universe was less than 1 Gyr old (see, e.g., \citealt{kmaw20, victoria20} and references therein), i.e., 
at $z\ga6$, and that these first sources played a crucial role in the reionization that appears to have ended at z$\sim$6 \citep{thebob01, fan06}. Such a timescale is consistent with 
observations of the Cosmic Microwave Background from the \emph{Planck} telescope, which use the Thomson optical scattering depth to infer the epoch of an instantaneous reionization 
at $7.0\la z \la 8.5$ \citep{planck16, planck18}. While admirable attempts have been made to find and characterize galaxies formed during the EoR through large imaging and 
spectroscopic campaigns performed in blank fields, as well as around massive clusters of galaxies to harness the power of gravitational lensing, observational 
constraints, especially from a spectroscopic perspective, still remain sparse. At the highest redshifts, such campaigns are just now gaining the power to seriously challenge 
theoretical models of galaxy formation. Unsurprisingly, these galaxy populations remain a key driver of future instrumentation and major missions such as the 
\emph{James Webb Space Telescope} (\emph{JWST}; \citealt{gardner06}), the \emph{Nancy Roman Grace Space Telescope} \citep{spergel15}, the Thirty Meter Telescope 
(TMT; \citealt{sanders13}), Giant Magellan Telescope (GMT; \citealt{johns12}), and the Extremely Large Telescope (ELT; \citealt{gilmozzi07}).


The identification of high-redshift star-forming galaxies is often done through the $n=2 \rightarrow n=1$ Ly$\alpha$ 
$\lambda$1216\AA\ transition of hydrogen (hereafter Ly$\alpha$). Ly$\alpha$ emission primarily originates from the recombination of ionized hydrogen around newly-formed stars 
that emit large amounts of ionizing radiation. Ly$\alpha$ emission can be extremely strong as it is intrinsically the brightest recombination line \citep{pp67} and is situated 
well into the rest-frame 
UV. As such, Ly$\alpha$ is, in principle, easily detected at very high redshifts, giving it large potential as a tool for high-redshift galaxy evolution and cosmological studies.
Other rest-frame UV lines such as CIV $\lambda$1549\AA, HeII $\lambda$1640\AA, and CIII] $\lambda$1907,1909\AA\ are seen relatively rarely, are generally too faint, and, when bright,  
likely originate from galaxy populations too biased to particular phases, such as during the presence of AGN, to serve as a reliable, general substitute 
for Ly$\alpha$ line emission (e.g., \citealt{paolo13, schmidt16, kimihiko18, dong19}, though see 
\citealt{stark15a, stark15b} for an alternative view). While considerable headway has been made in measuring the [OIII] 88$\mu$m and [CII] 
158$\mu$m emission lines in high-redshift ($z\ga4$) galaxies primarily from the ALMA or the (Extended) Very Large Array (VLA) (e.g., 
\citealt{ferkinhoff10, capak15, inoue16, LPentz16, marusa17, carniani17, dong20, mbetz20}), most detected galaxies are required to have a redshift from rest-frame UV spectroscopy prior
to their targeting, as blind surveys for these lines are expensive (though see, e.g., \citealt{smit18, hashimoto18b, wenjia20, federica21} for counterexamples). 

Ly$\alpha$ emission is strongest when the ionizing stars are numerous, young, and an appreciable amount of atomic gas remains either \emph{in situ} or in close proximity. 
However, Ly$\alpha$ photons scatter resonantly, which creates numerous complications for its ability to escape the emitting galaxy, the circumgalactic medium, and 
the surrounding neutral intergalactic medium (IGM, e.g., 
\citealt{peebles93, dijkstra14}). These complications are compounded immensely in presence of dust in the interstellar medium (ISM) of galaxies, which serves to absorb Ly$\alpha$ 
photons during the resonant scattering process (e.g., \citealt{dayal11, hayes11, paolo15})
and leads to an expected decline in the observed fraction of Ly$\alpha$-emitters (LAEs) at a fixed equivalent width ($EW$) limit 
with decreasing redshift. The reverse trend is expected at $z\ga6$ due to the presence of 
an increasingly neutral IGM, such as that during reionization the fraction of LAEs again at a fixed $EW$ limit should decrease with increasing redshift. 
The confluence of these two effects appears to lead 
to a picture where the fraction of LAEs increases from $z\sim0$ until the edge of the reionization era ($z\sim6$) and then begins to drop precipitously at higher redshifts. Such 
a trend is indeed observed when the fraction of LAEs or derivative Ly$\alpha$ quantities have been used to probe the physical conditions during the reionization epoch, with 
such studies pointing to evidence of an increasingly neutral medium at $z\ga6$ (e.g., \citealt{ota08, fontana10, LPentz11, LPentz18, ono12, treu13, schenker14, tilvi14, hu17, 
zheng17, austin19a, mason18, mason19, spencer20}, though see, e.g., \citealt{debarros17, caruana18, kusakabe20} for a slightly different view). The fractional amount of Ly$\alpha$ 
estimated to escape star-forming galaxies can also be used to place constraints on 
internal galaxy physics (e.g., \citealt{dijkstra11, hayes11, debarros17, sobral18}) and the measured velocity difference between the Ly$\alpha$ feature and the systemic velocity as well 
as the Ly$\alpha$ line profile can be used to constrain kinematic models of such galaxies (e.g., \citealt{dawson02, westra05, marcin08, verhamme08, verhamme15, dilettantedaniel08, 
steidel10, dijkstra14, loopylucia17, lillaura19, paolo20}). Additionally, various Ly$\alpha$ properties may be helpful in indicating the amount of ionizing photons escaping galaxies 
(e.g., \citealt{dijkstra16, verhamme17, lillaura18, steidel18, mason20}), a key element of reionization models.

While such properties make Ly$\alpha$ an extremely useful tool for probing the physics of high-redshift galaxies and of reionization, it is also an extremely 
frustrating one. Without the presence of Ly$\alpha$, or the other few emission lines that are detectable at high-redshift given observational and temporal constraints, 
the veracity of photometric redshift estimates cannot be determined. That Ly$\alpha$ does not appear in emission for the majority of galaxies at all redshifts blunts its effectiveness
as a tool to confirm redshifts. While mitigating measures involving the full photometric redshift probability density function, $P(z)$, in conjunction with other constraints 
can be used even when Ly$\alpha$ is not present (e.g., \citealt{mason18, austin19a, mason19, spencer20}), other, more subtle issues remain with the Ly$\alpha$ feature. 

Early investigations of the spatial size of Ly$\alpha$ emission using \emph{HST} data (e.g., \citealt{finkel11}), ground-based narrowband imaging (e.g., \citealt{steidel11}), and predictions 
from simulations (e.g., \citealt{zheng11}) all indicated that Ly$\alpha$ sizes might exceed those in the rest-frame UV. This has now been confirmed, as studies of large  
numbers of galaxies at $3\la z \la 6$ observed with
integral field unit (IFU) spectroscopy from the Keck Cosmic Web Imager (KCWI; \citealt{morrissey18}) and the Multi-Unit Spectroscopic Explorer (MUSE; \citealt{bacon10,bacon14}) 
have clearly demonstrated that, on average, Ly$\alpha$ halo sizes of LAEs well exceed their UV extent, with typical size ratios of $\sim$3-10 
(e.g., \citealt{wisotzki16,wisotzki18,leclercq17,erb18}). 
Such differences are not limited only to sizes; in local analogs of high-redshift star-forming galaxies, Ly$\alpha$ emission exibits differences from rest-frame UV 
emission in a variety of morphological measures \citep{loopylucia15}. Additionally, recent work on large 
samples of LAEs over a similar redshift range ($2\la z \la 5$) observed with slit spectroscopy from the Very Large Telescope array (VLT) have shown that galaxies can exhibit
non-negligible spatial offsets in their Ly$\alpha$ emission relative to that of their rest-frame UV continuum \citep{austin19b}, with large offsets ($\sim2$ proper kpc) 
observed in 10-20\% of star-forming galaxies near the characteristic UV luminosity, $\sim L^{\ast}_{UV}$, at $z\ga2$ \citep{bruno20}. These studies  
confirm earlier work on single galaxies 
or narrowband imaging observations of large samples of LAE candidates, of the existence of a class of LAE with large, significant Ly$\alpha$-UV spatial offsets 
\citep{bunker00, fynbo01, shibuya14}. Both the larger extent of the Ly$\alpha$ emitting region and the spatial offset from the rest-frame UV continuum 
location would lead to additional losses of Ly$\alpha$ flux
in observations relying on slit spectrographs, as slits are typically placed on the UV barycenter and slit widths are typically much smaller than the observed 
Ly$\alpha$ halo size. If Ly$\alpha$ halo sizes or spatial offsets were a strong function of redshift or of other physics related to the galaxy populations or their 
surrounding media, such effects could have severe impacts on the inference of galaxy properties and the evolution of the IGM during the EoR.

In this paper, we study the latter of these two effects, the projected Ly$\alpha$-UV spatial offset, in a sample of 36 high-redshift ($5 \la z \la 7$) lensed LAEs taken 
from \citet{spencer20} and 29 high-redshift non-lensed LAEs drawn from \citet{LPentz11} and \citet{LPentz18}. This represents the first such study 
performed on a large sample of galaxies at this redshift. Due in part to lensing, the combined galaxy sample spans 
more than three orders of magnitude in intrinsic UV brightness, which allows for a comparison of offsets in UV-brighter and UV-fainter samples. Additionally, we 
compare to the few samples available in the literature at $2\la z \la5$ for which this measurement has been made. The paper is structured as follows. In \S\ref{sec:data}
we present the samples used in this study, the associated data, and the process of measuring various quantities including the Ly$\alpha$-UV offset for both
our magnified and unmagnified sample. In \S\ref{results} we discuss the relationship between the Ly$\alpha$-UV offset and other properties of the galaxies including
redshift, discuss the possible genesis for these offsets, and quantify the amount of additional slit loss resulting from these offsets for a variety of observational
setups. \S\ref{conclusions} presents our conclusions.  

Throughout this paper all magnitudes are presented in the AB system \citep{okengunn83, fukugita96} and distances are given in proper rather than comoving units. 
We adopt a concordance $\Lambda$CDM cosmology with $H_{0}$ = 70 km s$^{-1}$ Mpc$^{-1}$, $\Omega_{\Lambda}$ = 0.73, and $\Omega_{M}$ = 0.27. While abbreviated for convenience, 
throughout the paper absolute magnitudes are in units of $M_{AB}+5\log(h_{70})$, luminosities in units of $h_{70}^{-2}$ ergs s$^{-1}$, and proper distances and volumes 
in units of $h_{70}^{-1}$ kpc and $h_{70}^{-3}$ Mpc$^{3}$, respectively, where $h_{70}\equiv H_{0}/70$ km$^{-1}$ s Mpc. All equivalent widths are given in the rest-frame.

\section{Data and Analysis} 

\label{sec:data}

In this section we present both sets of high-redshift LAEs used in this paper as well as the measurements of various quantities that enter 
into our analysis, which include the estimate of spatial offsets between the observed Ly$\alpha$ feature and the rest-frame UV continuum. 
It is important to note here that, since we are observing these galaxies with slit spectroscopy instead of, e.g., integral field unit spectroscopy, we can only constrain 
those offsets which occur along one-dimension, namely that parallel to the major axis of the slit. Thus, all offsets presented here are projected offsets
in this dimension implying they are necessarily lower limits, as is the frequency of significant offsets. While we use the term ``spatial offset" throughout the
paper, we ask the reader to keep in mind that we always refer to the projected offset.

\subsection{Magnified Sample}  \label{subsec:SURFSUP}

The majority of our magnified sample is drawn from the sample presented in \citet{spencer20}. This sample includes 36 LAEs\footnote{While not all galaxies in our
sample have $EW$s in excess of the traditional limit used to define LAEs (i.e., 20-25\AA), we adopt this term for simplicity.} observed 
over the redshift range $5\la z \la 7$, each lensed by a massive foreground cluster. These LAEs were compiled from a massive, $\sim$200 
hour\footnote{Note that this number refers to the total time spent on the telescope rather than the sum of the DEIMOS integration time, which is roughly 100 hours.} 		
spectroscopic campaign with the DEep Imaging Multi-Object Spectrograph (DEIMOS; \citealt{fab03}) situated at the right Nasmyth focus of the Keck {\small II} telescope.
In total, 10 cluster fields were targeted with Keck/DEIMOS. These clusters include four of the five \emph{Hubble} Frontier Fields (HFFs; \citealt{abell89, ebeling01, lotz17}), 
several clusters from the  
Cluster Lensing And Supernova survey with Hubble (CLASH; \citealt{postman12}), and other efficient lensing clusters with comparable data from the 
Spitzer UltRa Faint SUrvey Program (SURFSUP; \citealt{marusa14}). Each cluster was 
observed with the Advanced Camera for Surveys (ACS; \citealt{ford98}) and the Wide Field Camera 3 (WFC3; \citealt{mackenty08}) aboard the \emph{Hubble} Space Telescope (\emph{HST}) 
in a large complement of broadband filters to a depth of $\sim$27-29 per filter. In addition, all clusters were imaged with the non-cryogenic bands of the 
\emph{Spitzer} Space Telescope InfraRed Array Camera (IRAC; \citealt{fazio04}) to $3\sigma$ depths of $m_{AB}\sim26-27$ per filter \citep{wrustle14, marusa14, kuang16, lotz17}. These 
imaging data were used to select high-redshift candidate objects through a series of color-color 
and photometric redshift (hereafter $z_{phot}$) cuts for spectroscopic targeting with Keck/DEIMOS. In addition, potential high-redshift line emitters observed in 
the Grism Lens-Amplified Survey from Space (GLASS; \citealt{schmidt14, schmidt16, treu15}) survey were also selected and prioritized as targets. High-redshift candidates were 
targeted, in addition to arcs and potential cluster members (referred to broadly as ``filler slits''), with multi-object slitmasks to a depth ranging from  
an exposure time, $\tau_{exp}$ of 3480s-19200s across all masks ($\langle \tau_{exp} \rangle \sim 7800$s). Slit widths of 1$\arcsec$ were used for all observations.
Note that, in many cases, high-redshift candidates appeared on multiple masks (see \citealt{spencer20} for more details). In all such cases, the spectrum from each individual mask 
was used to constrain the various parameters used in this study and an error-weighted sum was used to obtain the final values for each candidate. 

\begin{figure*}
\plottwo{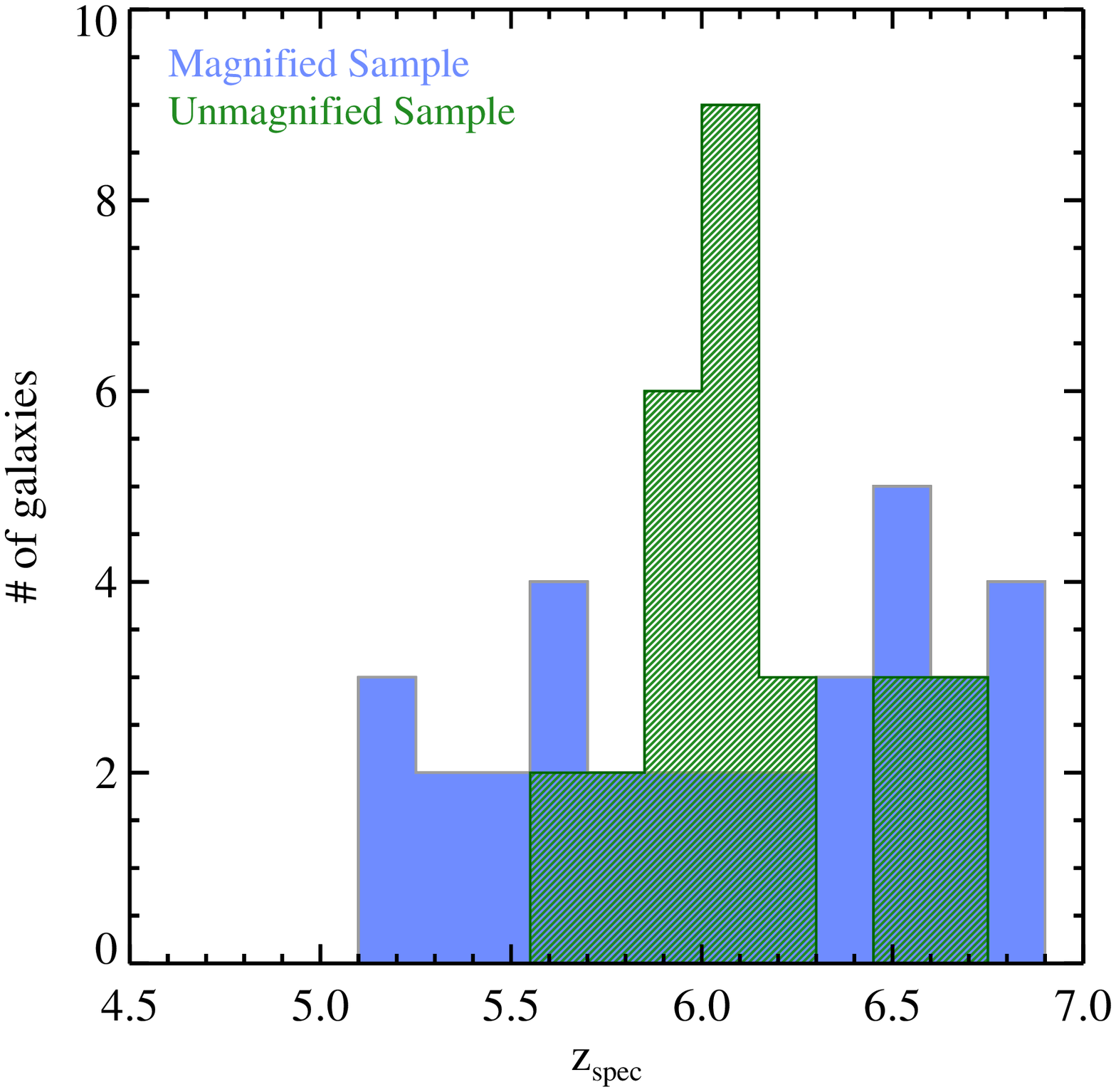}{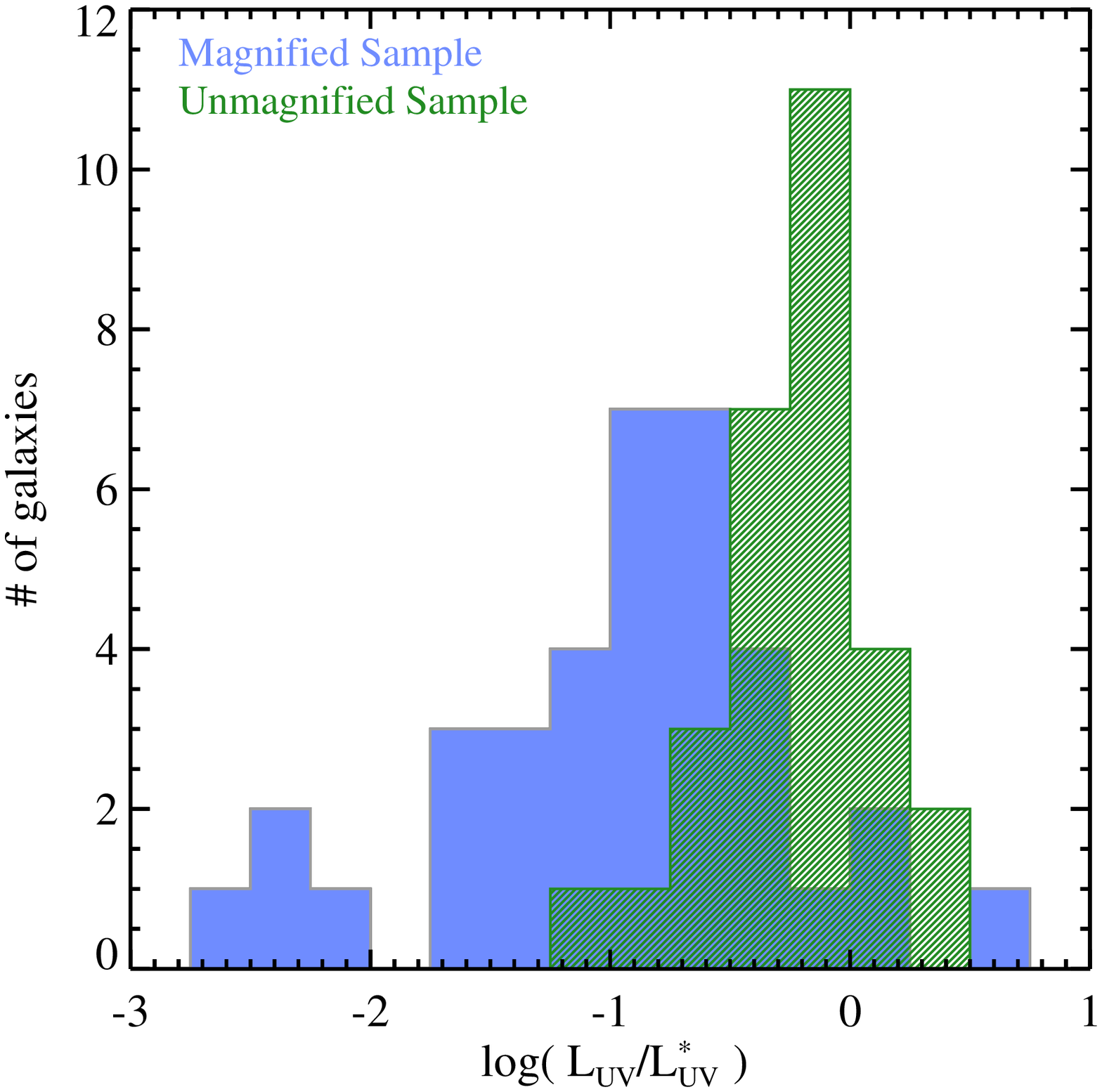}
\caption{\emph{Left}: Redshift distribution of our final lensed (blue solid histogram) and unlensed (green hatched histogram) sample of high-redshift
Ly$\alpha$ emitters. The two samples are demarcated at 0.25$L_{UV}/L^{\ast}_{UV}$ in order to create sub-samples of roughly equal size. One lower-redshift 
galaxy at $z_{spec}=4.4830$ taken from the sample of 
\citet{spencer20} is outside the bounds of this plot. \emph{Right:} Estimate of the rest-frame UV luminosity (magnification
corrected when applicable) relative to the characteristic UV luminosity at the redshift of each galaxy taken from \citet{anykey15}. The colors and
shading are identical to the left panel. The median UV luminosity of the combined sample is $\log(\widetilde{L_{UV}}/L^{\ast}_{UV})=-0.5$ and the entire
sample spans approximately three and a half orders of magnitude in $L_{UV}$.}
\label{fig:sampledist}
\end{figure*}

At the completion of the observing campaign, a final high-redshift candidate sample of 198 objects was compiled in \citet{spencer20}. This final sample 
relied on a uniform criterion involving 
the likelihood of the galaxy being in a redshift window that made the observation of the Ly$\alpha$ feature possible. Additionally, a systematic search was performed for 
high-redshift candidates that were serendipitously subtended by one or several of the observed slits. A semi-automated search of the one-dimensional signal-to-noise 
spectra of both targeted and serendipitous high-z candidates resulted in a final sample of 36 LAEs. All 36 of these LAEs had both visual 
emission line detections and formal line significances typically well in excess of 3$\sigma$. The median spectroscopic redshift of the 36 LAEs in our sample is $\tilde{z}_{spec}=6.23$ and the 
redshift distribution of these galaxies is shown in the left panel of Figure \ref{fig:sampledist}. A more complete account of the observations, reduction, 
and analysis of these data are given in \citet{spencer20}.  

\subsubsection{Ly$\alpha$ Spatial centroids of the Magnified Sample} \label{sec:Lyacent1}

To measure the Ly$\alpha$\footnote{While there remains some ambiguity as to the nature of the emission line, as discussed in \citet{spencer20}, the
possibility that these galaxies are genuine LAEs is very high. Therefore, we refer to this line simply as Ly$\alpha$ for the remainder of the paper.} location  
of the LAEs from \citet{spencer20} we collapsed the two-dimensional spectrum over the wavelength range where the Ly$\alpha$ emission appears in excess 
of the background. In the observed-frame this width was typically $\sim$8\AA. Note that this procedure was done for each individual night that a LAE was 
observed and the resultant values were combined for a given galaxy by a weighted average of all observations, where the signal-to-noise ratio (S/N) of the 
Ly$\alpha$ line in each observation is used as the weight. Prior to a formal measurement of the Ly$\alpha$ location, because the two-dimensional DEIMOS slits reduced through 
the \textsc{spec2d} package have an ambiguity in their directionality, which, in turn, results in an ambiguity in the cardinal direction of any observed offset, we 
measured the distance from the southern edge of the slit\footnote{No slits had sky PAs of 90$^{\circ}$ or 270$^{\circ}$, which meant the ``southern edge of the slit'' was always a defined 
quantity} to the rest-frame UV location of the target in each slit. This distance was then compared to the design location encoded by 
\textsc{dsimulator}\footnote{\url{https://www2.keck.hawaii.edu/inst/deimos/dsim.html}}, the DEIMOS mask design software (see \S\ref{sec:UVcent1} for more details on \textsc{dsimulator}). 
Coincident values meant the two-dimensional spectrum was oriented such that increasing pixel values along the major axis of the slit corresponded to more northerly  
movement\footnote{For slit PAs other than 0$^{\circ}$ and 180$^{\circ}$, movement along the major axis of the slit also results in movement to the east or west, the magnitude of
which is governed by the PA of a given slit. However, once the ambiguity between the directionality of the slit array in the north/south directions is set, the components of
the movement per spatial pixel in each of the cardinal directions becomes a trivial calculation.}, discordant values meant the slit was flipped, and increasing pixel values 
along the major axis of the slit resulted in movement southward.    

Once the spectrum was collapsed and the orientation determined, we used two methods to attempt to determine the spatial centroid
of the Ly$\alpha$ emission. The first and most common method used was to fit a Gaussian to the collapsed one-dimensional spatial profile of the Ly$\alpha$ emission
through a $\chi^2$-minimization technique using the \textsc{IDL}-based package \textsc{mpfitexpr}. The errors in each case were derived from the official \textsc{spec2d}
inverse-variance spectrum. The mean of the Gaussian fit was adopted as the raw\footnote{This is referred to as the ``raw'' value here as the spatial centroid
will be corrected for differential atmospheric refraction later in this section.} spatial centroid and the associated uncertainties were taken from the covariance matrix.
The second method, which was used only in a small percentage ($\sim$15\%) of cases where the S/N was generally too low or artifacts prevented a clean Gaussian fit, 
relied on the non-parametric approach of \citet{teague18}. Briefly, this approach identifies the pixel value where the collapsed spectrum is at maximum and 
uses this value, along with the value of the two directly adjacent pixels, to estimate the curvature of the spatial profile on either side of the maximum. A quadratic 
function then uses this input to estimate the line centroid. In essentially all high S/N cases where spectra were devoid of strong artifacts, the two methodologies 
returned similar results within the errors. 



Because of the ability of differential atmospheric refraction (DAR) to alter the spatial location at which an object is seen at different wavelengths, the measurement of 
Ly$\alpha$ location must either occur in close spectral proximity to the rest-frame ultraviolet (UV) continuum measurement (i.e., similar wavelengths) or a 
correction must be applied to account for this effect. In the case where the continuum is present in the spectra, this issue is obviated with a 
continuum measurement just redward of the Ly$\alpha$ feature, as indeed was done for our unmagnified sample (see \S\ref{subsec:LPentz}). In the case
of our magnified sample, however, little to no continuum is observed, requiring a correction for this effect. In the next section, we will estimate 
the spatial location of the UV continuum at the central wavelength of each mask (8800\AA\ for the vast majority of our data). As such, we corrected the 
raw Ly$\alpha$ spatial location to that expected at the central wavelength of the mask it is observed on using the method described in Appendix \ref{appendix:A}.  

\subsubsection{Rest-frame UV Spatial centroids of the Magnified Sample} \label{sec:UVcent1}

Because of the absence of the continuum in the spectral observations of the magnified sample, 
we are required to estimate the rest-frame UV location of these galaxies based on its expected location from \textsc{dsimulator}. 
Note that a few galaxies, especially those which serendipitously fell on the slit (see Appendix \ref{appendix:A}), were not centered along the minor axis 
of the slit and, thus, the expected UV position was also a projected one. This expected location is based on the right ascension and declination 
($\alpha$,$\delta$) of each target and corrected for the effect of 
DAR \emph{at the central wavelength of a given mask} using the expected airmass and position angle
(PA) relative to the parallactic angle of the observation, values which are input by the user at the time of the mask design. Because the actual observations are 
taken at a variety of airmass and PAs, typically over the course of many nights, the true DAR correction will be different than the one expected from the values
input during the mask design process. In Appendix \ref{appendix:A} we quantify the magnitude of this effect and the estimate of the UV centroid for 
LAE candidates that were serendipitously subtended by our slits. Additionally, we also discuss in Appendix \ref{appendix:A} potential scatter in our measurements
due to relative astrometric errors and conclude that such uncertainties are small enough to be ignored. 

At the end of this process described in Appendix \ref{appendix:A}, we have estimates of the rest-frame UV and Ly$\alpha$ pixel locations corrected for DAR 
and mask design issues. These two estimates are subtracted, the absolute value is taken, and the resultant number is multiplied by the DEIMOS spatial plate scale (0.1185$\arcsec$ pix$^{-1}$) 
and the angular scale at the redshift of the LAE to get the main quantity used in this analysis: $|\Delta_{\rm{Ly}\alpha-\rm{UV}}|$ in units of proper kpc (pkpc). Since
all galaxies in this sample are lensed, we make the simple assumption here that the galaxy is enlarged with azimuthal symmetry and the true distance measured is given as 
$|\Delta_{\rm{Ly}\alpha-\rm{UV}}|\equiv|\Delta_{\rm{Ly}\alpha-\rm{UV}}|_{lensed}/\sqrt{\mu}$, where $\mu$ is the magnification of each galaxy as estimated by the formalism
of \citet{marusa05,marusa09}. For more detail on the lensing reconstruction for the fields studied here see \citet{austin19a} and references therein. In Figure \ref{fig:offsetexample} 
we show examples drawn from our full sample of strong Ly$\alpha$ emission offset at both a typical and extreme level from the UV continuum emission. 

The vast majority of the galaxies in our sample have relatively small magnification factors ($\mu\la5$). Therefore, the validity of the above assumption is of limited consequences 
for our results and, indeed, our main results remain unchanged if we do not assume the magnification is azimuthally symmetric and, instead, assume a variety of different ellipticities.
This logic applies as well to our ignorance of the true magnification, an ignorance which is present both within the confines of a given model as well as the variation in the recovered
values for different modeling approaches. For the former, the formal uncertainties within our own models of the estimated $\mu$ values, as estimated from the interquartile range 
of bootstrap-resampled maps (see \citealt{austin19a} for more details), are small enough to be ignorable for the vast majority of galaxies. Specifically, the median uncertainty resulting 
from this process is 3\%, with $<5$\% of our sample showing magnification uncertainties $>$15\%, a negligible uncertainty considering the $\sqrt{\mu}$ dependence of our results. For the 
latter, the variation in $\mu$ as derived by different modeling approaches, we performed the following test. For the $\sim$40\% of our galaxies that lie within the HFFs, we calculated the 
3$\sigma$ clipped mean of the $\mu$ values for a large variety of models provided on a web-based tool specifically designed to compare the results of different 
models\footnote{\url{https://archive.stsci.edu/prepds/frontier/lensmodels/webtool}} and compared these to the $\mu$ values coming from our models. While individual estimates 
had sigma-clipped dispersions of factors of $\sim$2-3, the values of $\mu$ coming from our models were, on average, statistically consistent with the aggregate mean. Thus, while 
lensing-corrected $|\Delta_{\rm{Ly}\alpha-\rm{UV}}|$ values for individual galaxies likely contain an additional uncertainty factor of $\sqrt{2}-\sqrt{3}$ due to our 
ignorance of the true $\mu$ value, this uncertainty likely does not affect the average values that the main conclusions of this paper are based on. 

\subsection{Unmagnified Sample}  \label{subsec:LPentz}

The second sample of galaxies used in this work, which we will broadly refer to as the ``unmagnified sample'' as these observations are taken in legacy fields 
mostly devoid of massive, lower-redshift structure, was drawn from the samples of \citet{LPentz11} and \citet{LPentz18}, with a few additional galaxies 
taken from \citet{caruana14}. These galaxies were observed with FOcal Reducer/low dispersion Spectrograph 2 (FORS2; \citealt{app98}) 
equipped on the Very Large Telescope (VLT) Antu Unit Telescope and selected 
for spectroscopy through a variety of dropout and color criteria. Except for a few galaxies contained in the New Technology Telescope Deep Field 
(NTTDF; \citealt{arnouts99, fontana00}) and the BDF-4 field \citep{lehnert03, castellano10}, all galaxies in this sample were observed with deep multi-band \emph{HST}/ACS and WFC3
imaging as part of the Cosmic Assembly Near-infrared Deep Extragalactic Legacy Survey (CANDELS; \citealt{grogin11, koekemoer11}). In all cases, slit widths were set
to 1$\arcsec$. For more details on the imaging and spectral observations of this sample, see \citet{LPentz18} and references therein. 

We also considered the inclusion of additional galaxies drawing from the $5\le z\le 6$ presented in \citet{yana20} to supplement the unmagnified sample. This sample, which is 
comprised both of LAEs and non-emitters observed with extremely deep low-resolution spectra from the Visible Imaging Multi-Object Spectrograph (VIMOS; \citealt{dong03}) previously 
mounted on the VLT Melipal Unit Telescope, represents the high-redshift tail of galaxies in the VIMOS Ultra-Deep Survey (VUDS; \citealt{dong15}). However, despite the 
extremely deep spectroscopy, we were
only able to reliably measure $|\Delta_{\rm{Ly}\alpha-\rm{UV}}|$ for a sub-sample of six LAEs due primarily to lack of observed continuum and uncertainty 
in the expected UV location of the target in the reduced data. While these six galaxies are included as a comparison to our data in \S\ref{LyaUVoff}, due to the relatively 
small number of galaxies and the differing selection, observational strategy, and methods used to measure associated quantities for these galaxies, we chose not to include 
these galaxies in our final sample. We note, however, that none of the results presented in this study are meaningfully changed if we instead included these galaxies as part of the 
unmagnified sample. 

\subsubsection{Ly$\alpha$ and Rest-frame UV Spatial Centroids of the Unmagnified Sample}
\label{LPentzmeasure}
For this sample, the continuum was observed for a large number of galaxies, and, as such, a more direct approach could be taken to measure the offset between Ly$\alpha$
and the continuum. For each of the $\sim$50 galaxies in the sample, the Ly$\alpha$ location was measured using the methods given in section \ref{sec:Lyacent1} with the 
Gaussian method always preferred except in low-S/N cases. In the absence of formal error spectra, errors were assumed to be Poissonian. Though such 
an assumption likely underestimates the true noise, the results did not meaningfully change if we instead scale the errors to match the average root-mean-square fluctuations. 
To measure the UV centroid, the flux over the same spatial window was summed over a broad spectral window at least 10\AA\ in the observed-frame
redward of the observed Ly$\alpha$ line. The spectral window employed changed based on the target to avoid regions of dominant skylines and those regions where the continuum
signal was minimal, but was typically several 100\AA\ (observed-frame) wide. The reduction process of the FORS2 data employed spatial trimming of the data in order to 
maximize the effectiveness of the sky subtraction. At the completion of this process, the information on the expected location of the galaxies was no longer encoded 
in the data, and, as such, only galaxies with well-measured continuum and Ly$\alpha$ centroids were allowed to remain in our sample. This cut resulted in a total of 29 
galaxies. The $|\Delta_{\rm{Ly}\alpha-\rm{UV}}|$ was then calculated in a similar manner to those in magnified sample using the appropriate plate
scale, though, since Ly$\alpha$ and continuum centroids were measured at similar wavelengths, no DAR correction was applied to either value. Additionally, since galaxies in this sample 
are, at least to first approximation, not lensed modulo small weak-lensing effects that may occur along a random line of sight, no lensing correction to the measured 
offset is applied. The median redshift of the 29 galaxies in this sample is $\tilde{z}_{spec}$=6.05
and the redshift distribution is shown in the left panel of Figure \ref{fig:sampledist}. 

Because we are using different approaches to estimate $|\Delta_{\rm{Ly}\alpha-\rm{UV}}|$ in the lensed and unlensed sample, we compared the $|\Delta_{\rm{Ly}\alpha-\rm{UV}}|$
distribution for the two samples where there was overlap in intrinsic UV luminosity. No statistically significant difference was found from a Kolmogorov-Smirnov (K-S) test.
Additionally, a Spearman's test finds no evidence of significant correlation between $|\Delta_{\rm{Ly}\alpha-\rm{UV}}|$ and the S/N of either the Ly$\alpha$ detection or 
the UV continuum in the spectroscopy (the latter could be tested only in the case of the unlensed sample). These tests indicate that any observed Ly$\alpha$-UV offset 
is likely to be real and not the result of poor centroiding. 
 
\subsection{Associated Quantities} \label{sec:LyaEWnstuff}

In \S\ref{LyaUVoff} we will explore whether $|\Delta_{\rm{Ly}\alpha-\rm{UV}}|$ depends on several basic properties of the galaxies as estimated through the various 
spectroscopic and imaging observations. These properties are the spectral redshift, the intrinsic (i.e., lensing corrected) far-UV luminosity relative to the characteristic
equivalent luminosity at the redshift of the galaxy ($L_{UV}/L^{\ast}_{UV}$), the intrinsic Ly$\alpha$ line luminosity ($L_{\rm{Ly}\alpha}$), and the rest-frame equivalent width 
of the Ly$\alpha$ line ($EW_{\rm{Ly}\alpha}$). We quickly review how these quantities were estimated for our data. 

For the magnified sample, $L_{UV}$ is estimated using the apparent magnitude of each LAE in the $F160W$ band, with a $k-$correction applied to convert
to rest-frame $\sim$1600\AA, and we convert to an intrinsic value by dividing the $k$-corrected $L_{UV}$ by the most likely magnification value, $\mu$. 
The redshift-dependent characteristic UV luminosity, $L^{\ast}_{UV}$, 
is taken from \citet{anykey15}. The $L_{\rm{Ly}\alpha}$ for each galaxy is estimated from the observed line flux corrected for slitloss effects assuming that the UV location and 
the Ly$\alpha$ line are co-located with respect to the slit. These values are also corrected for the magnification. The $EW_{\rm{Ly}\alpha}$ values are calculated 
using the de-magnified line luminosity and the rest-frame UV flux density was estimated from a filter whose throughput begins just redward of the observed 
wavelength of Ly$\alpha$ (typically $F105W$). For more details on these calculations see \citet{spencer20}.

For the unmagnified sample, equivalent values were drawn directly from the relevant references when available. In the case of \citet{LPentz11, LPentz18}, 
rest-frame absolute magnitudes ($M_{UV}$) were calculated in a similar manner to those in our magnified sample and were subsequently converted to 
$L_{UV}/L^{\ast}_{UV}$ values again using the $L^{\ast}_{UV}$ of \citet{anykey15}. In those works, slit losses were assumed to be small ($\sim$15\%) and were 
not included in estimating Ly$\alpha$ line luminosities and $EW$s. The $EW_{\rm{Ly}\alpha}$ values reported in those papers also relied on flux density measured 
in the broadband photometry in order to quantify the strength of the continuum redward of Lya. Additional data from the literature 
had their equivalent quantities derived in a similar manner. For more details, see \citet{LPentz18}. In the 
right panel of Figure \ref{fig:sampledist} we show the $L_{UV}/L^{\ast}_{UV}$ distribution for both the magnified and unmagnified sample. 


\begin{figure*}
\plottwospecial{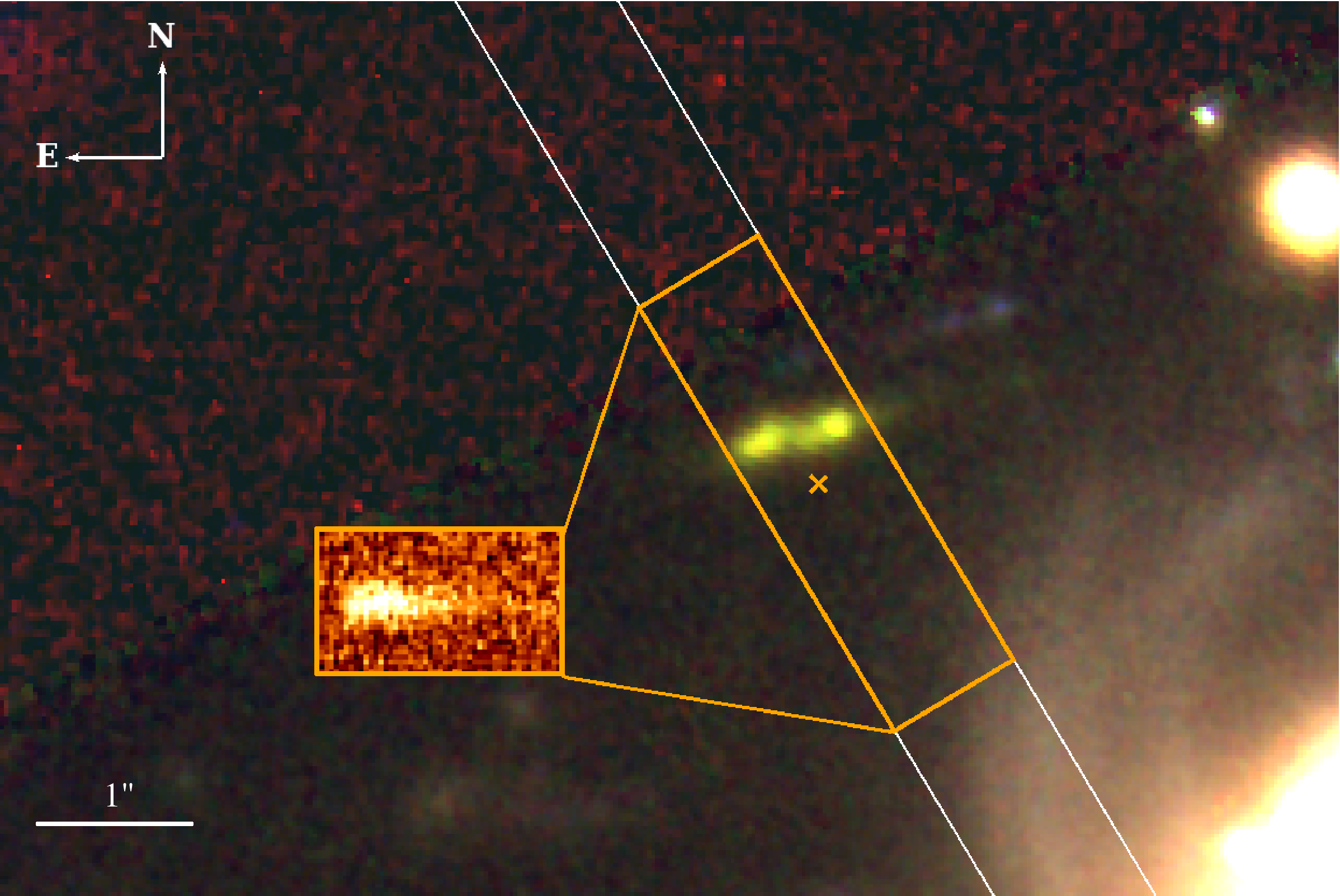}{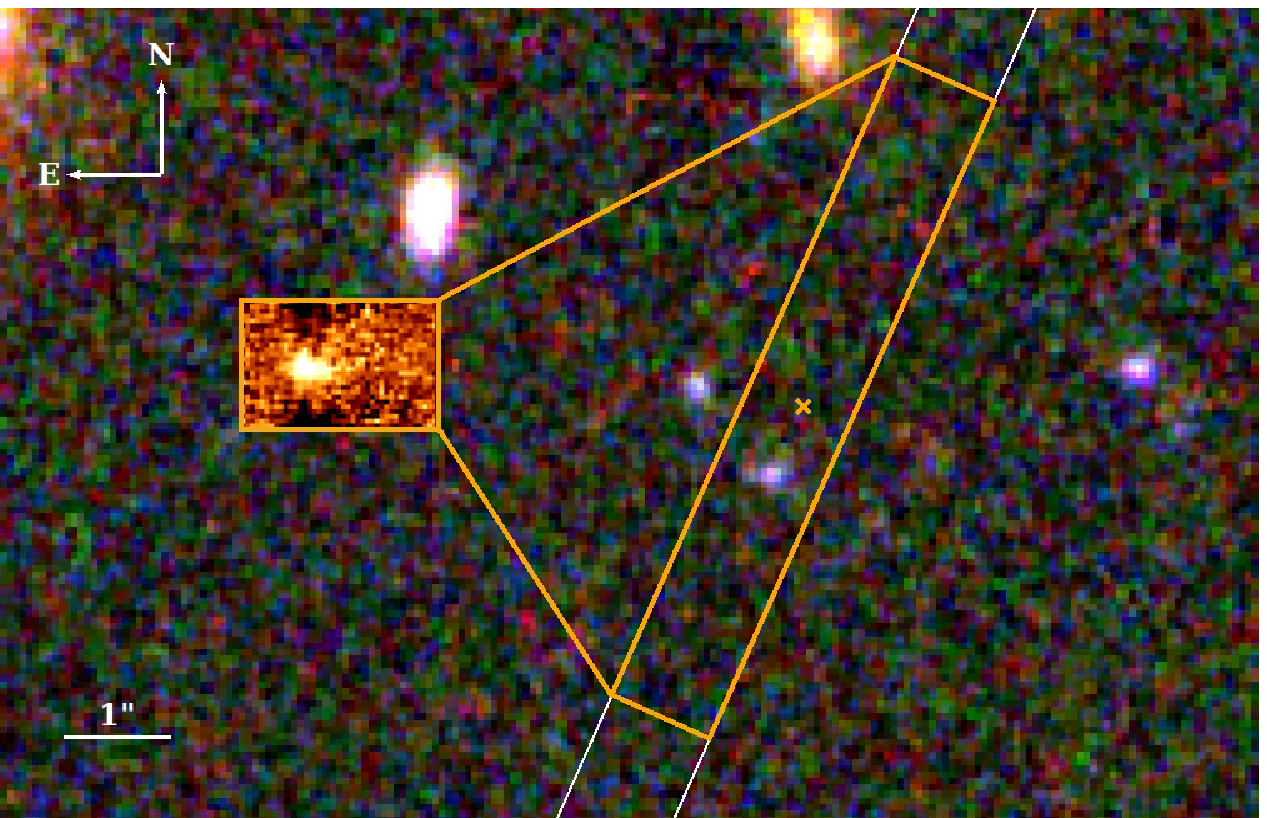}
\caption{\emph{Left:} Example of a magnified galaxy (ID=370.14 in \citealt{spencer20}) with an offset of $|\Delta_{\rm{Ly}\alpha-\rm{UV}}|=0.67$ kpc 
after correcting for lensing effects ($\mu=8.94$). This offset is typical for our full sample of LAEs, though on the large end for a lower-luminosity galaxy 
($L_{UV}/L^{\ast}_{UV}=0.22$). The background image is a $F814W/F105W/F160W$ color composite. The slit used to observe this galaxy is outlined in white. 
A portion of the two-dimensional spectrum corresponding to the spatial region highlighted in orange and the spectral region of the Ly$\alpha$ line is shown in the inset. 
Note that the two-dimensional spectrum inset is rotated and rectified for clarity such that the spatial axis, i.e., the direction along the major axis of the slit, 
is vertical. The observed spatial centroid of the Ly$\alpha$ line is shown as the orange $\times$. Note that while this galaxy exhibits multiple components possibly indicative
of a companion or multiple star-forming clumps, the direction of the Ly$\alpha$ offset is perpendicular to the position angle of these components. For reference, a 
scale bar indicating 1$\arcsec$ is shown in the bottom left. \emph{Right:} Similar to the left panel, but for the unmagnified galaxy UDS-23719 taken from 
\citet{LPentz18}. An $F125W$ image is used in the green channel as a substitute to $F105W$. This galaxy exhibits the largest offset in the full sample, with 
$|\Delta_{\rm{Ly}\alpha-\rm{UV}}|=3.65$ kpc.} 
\label{fig:offsetexample}
\end{figure*}

\subsection{Rest-frame UV Morphology and Galaxy Sizes}
\label{restframeUVmorph}

For all galaxies in both samples, save those few galaxies in the NTTDF and BDF-4 fields for which such imaging was not available, we visually inspected a \emph{HST}/WFC3 
$F160W$ postage stamp. This inspection was done to look for multiple clumps or other obvious features that may suggest that any appreciable Ly$\alpha$-UV offset is caused by 
merging activity or issues with centroiding the emitting galaxy in the UV. The observed $F160W$ band corresponds to rest-frame $\sim$2250\AA\ for the median redshift of the 
combined sample ($\tilde{z}_{spec}=6.15$). In the majority of cases ($\sim$65\%), the galaxy was observed to be an isolated galaxy with a reasonably symmetric 
surface brightness profile. The remaining $\sim$35\% appeared to be multicomponent systems, but the vast majority had either a slit orientation that was nearly perpendicular 
to the position angle of the multiple clumps that comprised the galaxy (see, e.g., the left panel of Figure \ref{fig:offsetexample}) or the directionality of the measured 
Ly$\alpha$-UV offset was opposite that of the direction to
the secondary clump(s)\footnote{Images for the magnified sample and the vast majority of the unmagnified sample are publicly available and slit orientations are shown
in \citet{LPentz18,spencer20} (for the former, slit PAs are identical those of the masks).}. We therefore conclude that the $|\Delta_{\rm{Ly}\alpha-\rm{UV}}|$ values measured in our sample are primarily the result of true 
\emph{intra}galaxy astrophysical offsets and do not result from misidentification of the original source or bad UV centroiding due to multiple components. 

In addition to qualitatively characterizing the morphology and multiplicity of our LAE samples, we also measured effective radii ($r_e$) using a combination of	
\textsc{Source Extractor} (\textsc{SExtractor}; \citealt{bertin96}) and \textsc{psfex} \citep{bertin11} to perform point-spread function (PSF)-corrected S\'{e}rsic fits on all 
galaxies imaged in \emph{HST}/WFC3 $F160W$. We chose to focus only on those galaxies with \emph{HST} imaging, which included all galaxies in the magnified and sample and 24 
galaxies in the unmagnified sample, with the S\'{e}rsic index treated as a free parameter. Extremely low S/N or visual reduction artifacts in the $F160W$ imaging 
caused the loss of $\sim$30\% of the galaxies in the magnified sample and one 
additional galaxy in the unmagnified sample. As in \S\ref{sec:UVcent1}, for magnified galaxies, measured $r_{e}$ were corrected to intrinsic by dividing by $\sqrt{\mu}$. 
Galaxies with multiple components were measured using \textsc{SExtractor} parameters specifically tuned to blend the multiple components
and perform the S\'{e}rsic fit on the composite system. This process resulted in a final sample of 48 galaxies with reliable $r_{e}$ measurements with an average 
$\tilde{r}_e = 0.6$ kpc, consistent with parametric measurements of other galaxy samples at similar redshifts (see, e.g., \citealt{shibuya15, shibuya19,
cl16,femalebruno18}). In later sections, we will compare these UV sizes with their respective Ly$\alpha$-UV offsets. All UV sizes, Ly$\alpha$-UV offsets, and all 
other properties of galaxies in the unmagnified and magnified samples are given in Tables \ref{tab:prop1} and \ref{tab:prop2}, respectively. 

\begin{figure*}
\plottwoalmostspecial{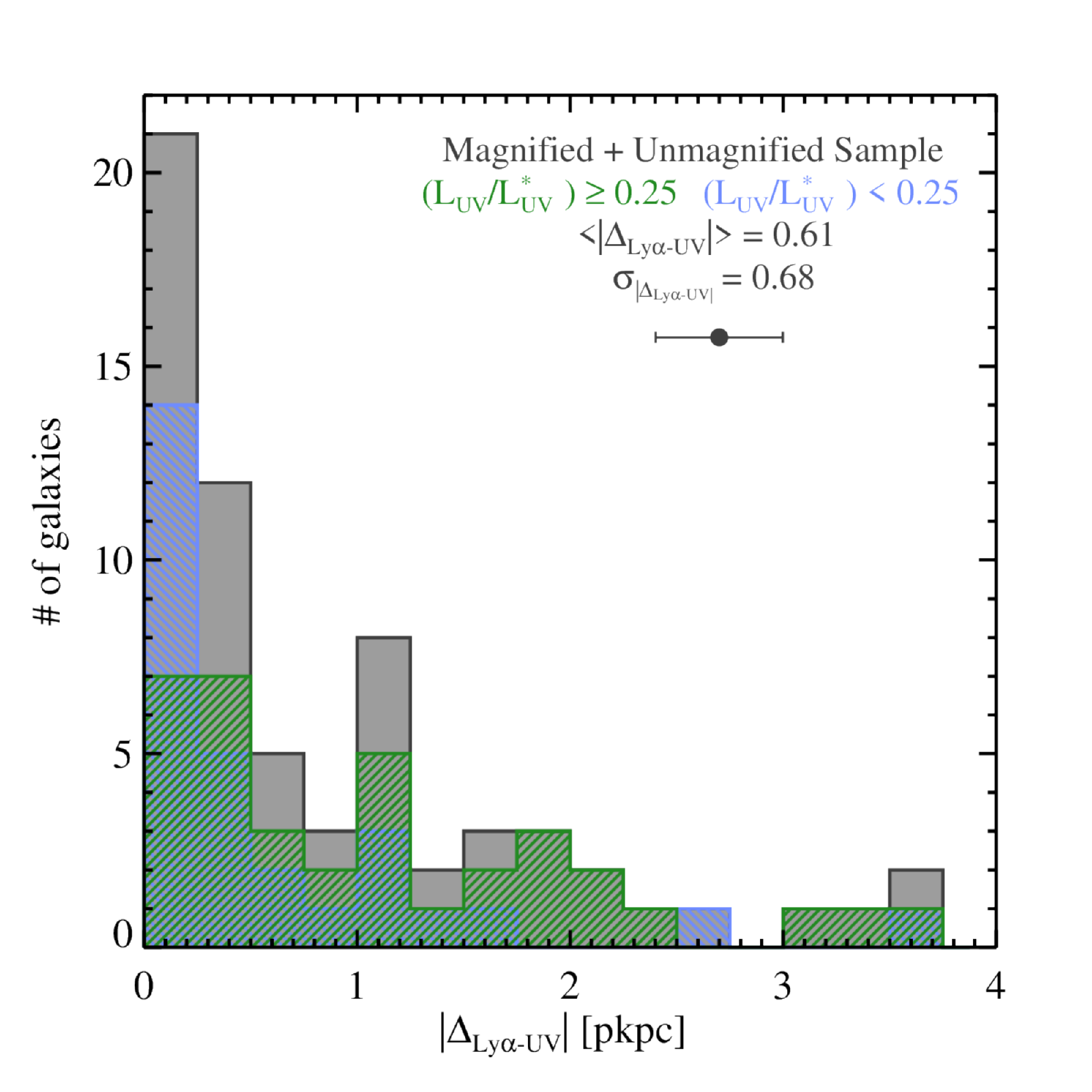}{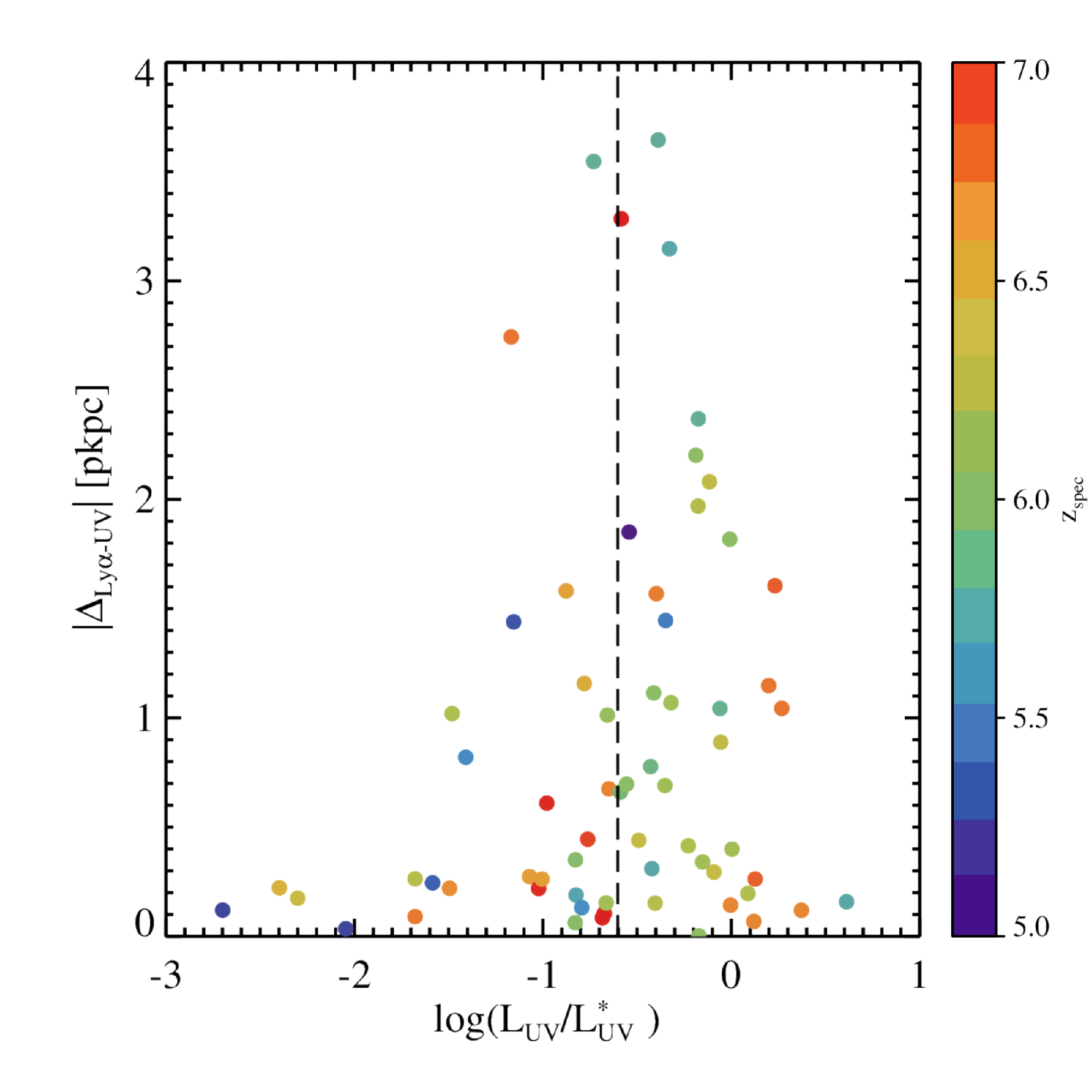}
\caption{\emph{Left:} Distribution of the Ly$\alpha$ and UV continuum spatial offset for our full (gray solid), fainter (light blue hatched),
and brighter (green hatched) samples. The definition of the two luminosity samples along with the median offset and NMAD scatter of the full
sample is given in the top right. The average measurement uncertainty for the combined sample is also shown. \emph{Right:} Ly$\alpha$-UV spatial offset plotted
against $L_{UV}/L^{\ast}_{UV}$ for our full sample. The color bar
indicates the spectroscopic redshift of each galaxy. The dashed line shows the delineation point between our lower- and higher-luminosity samples.
There is a marginally significant positive correlation observed between $|\Delta_{\rm{Ly}\alpha-\rm{UV}}|$
and $L_{UV}/L^{\ast}_{UV}$, and the UV-brighter sample shows both a larger percentage of significant offsets and a larger median offset than their
lower-luminosity counterparts (0.89$\pm$0.18 vs.\ 0.27$\pm$0.05 kpc, respectively). As discussed in \S\ref{LyaUVoff}, this difference is likely not due to
observational effects and rather represents real differences between the two galaxy populations. No significant correlation is observed as a function
of redshift within this sample.}
\label{fig:offsetfinal}
\end{figure*}

\begin{table*}
    \begin{center}
    \caption{Properties of the Unmagnified LAEs}			
    \label{tab:prop1}
    {\vskip 1mm}
    \begin{tabular}{c @{\hskip 15mm} c}

        \begin{tabular}{lccccccccc}

        \hline \\[-3.3mm]
        ID & RA & Dec & $z_{spec}$ & $|\Delta_{\rm{Ly}\alpha-\rm{UV}}|$ & $r_e$ & $L_{UV}/L^{\ast}_{UV}$ & $\log(L_{\rm{Ly}\alpha})$ & $EW$(Ly$\alpha$)$_{rest}$  \\[0mm]
         & & & & [pkpc] & [pkpc] & & [ergs s$^{-1}$] & [\AA] \\[0mm]
        \hline \\[-3.3mm]
        GOODS-10219 & 53.2020264 & -27.816353 & 6.136 & 2.0$\pm$0.4 & 1.2$\pm$0.6 & 0.67 & 42.3 & 14 \\[1mm]
        GOODS-11464 & 53.1174545 & -27.805187 & 5.939 & 2.2$\pm$0.4 & 0.5$\pm$0.2 & 0.65 & 42.5 & 22 \\[1mm]
        GOODS-14439 & 53.1248894 & -27.784107 & 5.783 & 0.8$\pm$0.2 & 1.6$\pm$0.1 & 0.37 & 42.6 & 50 \\[1mm]
        GOODS-15443 & 53.1519432 & -27.778177 & 5.938 & 1.8$\pm$0.7& 0.9$^{+2.5}_{-0.9}$ & 0.98 & 42.2 & 7 &\\[1mm]
        GOODS-16371 & 53.1595078 & -27.771446 & 6.108 & 0.4$\pm$0.7 & 0.6$\pm$0.5 & 0.59 & 42.1 & 11 \\[1mm]
        GOODS-17692 & 53.1627693 & -27.760759 & 5.916 & 0.1$\pm$0.4 & 0.3$\pm$1.2 & 0.15 & 43.0 & 153 \\[1mm]
        GOODS-17720 & 53.1225586 & -27.760504 & 5.927 & 1.1$\pm$0.9 & 1.0$\pm$0.7 & 0.39 & 42.4 & 30 \\[1mm]
        GOODS-18310 & 53.1419334 & -27.755154 & 6.046 & 0.2$\pm$0.5 & 0.5$\pm$0.2 & 0.22 & 42.1 & 26 \\[1mm]
        GOODS-31891 & 53.1742516 & -27.769789 & 6.630 & 2.7$\pm$0.7 & ---$^{a}$ & 0.07 & 42.0 & 73 \\[1mm]
        GOODS-32453 & 53.1799088 & -27.754873 & 5.929 & 0.7$\pm$0.3 & 0.6$^{+1.4}_{-0.6}$ & 0.28 & 42.5 & 56 \\[1mm]
        GOODS-33418 & 53.1944962 & -27.726349 & 7.058 & 3.3$\pm$0.3 & 0.2$\pm$0.2 & 0.26 & 42.9 & 110 \\[1mm]
        UDS-1920 & 34.4887581 & -5.2656999 & 6.565 & 0.1$\pm$0.3 & 0.9$\pm$0.4 & 2.35 & 42.2 & 3 \\[1mm]
        UDS-4812 & 34.4757347 & -5.2484999 & 6.561 & 0.3$\pm$0.1& 0.5$^{+0.6}_{-0.5}$ & 0.99 & 43.0 & 44 \\[1mm]
        UDS-4872 & 34.4820328 & -5.2481742 & 6.564 & 0.1$\pm$0.3 & 0.7$\pm$0.3 & 1.32 & 42.4 & 10 \\[1mm]
        UDS-14549 & 34.4828377 & -5.1953101 & 6.033 & 0.7$\pm$0.3 & 0.3$^{+0.4}_{-0.3}$ & 0.44 & 42.3 & 20 \\[1mm]
        UDS-15559 & 34.2315788 & -5.1897931 & 6.044 & 0.4$\pm$0.2 & 0.4$^{+0.7}_{-0.4}$ & 1.01 & 42.5 & 79 \\[1mm]
        UDS-18087 & 34.3972206 & -5.1756892 & 6.119 & 0.2$\pm$0.2 & 0.4$^{+0.5}_{-0.4}$ & 1.22 & 43.2 & 47 \\[1mm]
        UDS-23719 & 34.3104324 & -5.1456208 & 5.683 & 3.6$\pm$ 0.2& 0.7$^{+1.4}_{-0.7}$ & 0.41 & 42.6 & 64 \\[1mm]
        UDS-28306 & 34.3560867 & -5.2582278 & 6.142 & 0.2$\pm$0.1 & 1.1$\pm$0.7 & 0.39 & 42.5 & 41 \\[1mm]
        UDS-29191 & 34.5253906 & -5.2412128 & 5.943 & 0.0$\pm$0.3& 0.3$^{+0.5}_{-0.3}$ & 0.67 & 42.6 & 34 \\[1mm]
        COSMOS-7692 & 150.107757 & 2.2719180 & 6.046 & 0.1$\pm$ 0.1& 1.9$\pm$0.2 & 0.48 & 42.5 & 24 \\[1mm]
        COSMOS-21411 & 150.183018 & 2.4392050 & 6.221 &0.3$\pm$0.3 & 1.0$\pm$0.6 & 0.81 & 43.2 & 90 \\[1mm]
        COSMOS-24108 & 150.197222 & 2.4786510 & 6.629 & 1.0$\pm$0.3 & 1.6$\pm$1.3 & 1.85 & 43.0 & 27 \\[1mm]
        NTT-6345 & 181.403901 & -7.7561900 & 6.701 & 1.6$\pm$0.3 &  ---$^{a}$ & 1.70 & 41.7 & 15 \\[1mm]
        NTT-7246 & 181.380139 & -7.7700590 & 5.724 & 1.0$\pm$0.4 &  ---$^{a}$ & 0.87 & 42.2 & 12 \\[1mm]
        BDF-3995 & 336.979035 & -35.179460 & 6.198 & 2.1$\pm$0.6 &  ---$^{a}$ & 0.77 & 42.0 & 7 \\[1mm]
        BDF-4085 & 336.957328 & -35.181049 & 6.196 & 0.9$\pm$0.2 &  ---$^{a}$ & 0.88 & 43.0 & 110 \\[1mm]
        BDF-5870 & 336.955504 & -35.208964 & 5.632 & 3.1$\pm$0.3 & ---$^{a}$ & 0.47 & 41.8 & 10 \\[1mm]
        \hline \\[-3mm]
        \end{tabular}

    \end{tabular}

    \end{center}
$^a$ No \emph{HST}/WFC3 $F160W$ imaging or insufficient S/N
\end{table*}

\begin{table*}
    \begin{center}
    \caption{Properties of the Magnified LAEs}
    \label{tab:prop2}
    {\vskip 1mm}
    \begin{tabular}{c @{\hskip 15mm} c}

        \begin{tabular}{lccccccccc}

        \hline \\[-3.3mm]
        ID & RA & Dec & $z_{spec}$ & $|\Delta_{\rm{Ly}\alpha-\rm{UV}}|$$^{a}$ & $r_e$$^{a}$ & $L_{UV}/L^{\ast}_{UV}$$^{a}$ & $\log(L_{\rm{Ly}\alpha})^{a}$ & $EW$(Ly$\alpha$)$_{rest}$ & $\mu_{med}$ \\[0mm]
         & & & & [pkpc] & [pkpc] & & [ergs s$^{-1}$] & [\AA] \\[0mm]
        \hline \\[-3.3mm]
        2744.116 & 3.592285 & -30.409911 & 6.101 & 1.0$\pm$2.0 & 1.5$^{+2.1}_{-1.5}$  & 0.03 & 41.5 & 223 & 4.3 \\[1mm]
        370.14 & 39.978069 & -1.558956 & 6.572 & 0.7$\pm$0.1 & 0.8$\pm$0.4 & 0.22 & 42.0 & 80 & 8.9 \\[1mm]
        370.43 & 39.966302 & -1.587095 & 6.612 & 0.1$\pm$1.2 & 1.3$\pm$0.2 & 0.02 & 40.8 & 97 & 8.4 \\[1mm]
        370.55 & 39.990193 & -1.571209 & 5.614 & 0.2$\pm$0.1 & 0.3$\pm$0.3 & 0.15 & 41.9 & 101 & 3.2 \\[1mm]
        0416.17 & 64.036510 & -24.092300 & 5.993 & 1.0$\pm$0.2 & ---$^{b}$ & 0.22 & 41.9 & 66 & 2.6 \\[1mm]
        0416.56 & 64.047848 & -24.062069 & 6.146 & 0.3$\pm$0.3 & 0.3$\pm$0.1 & 0.02 & 40.8 & 105 & 22.6 \\[1mm]
        0416.89 & 64.048668 & -24.082184 & 5.241 & 1.4$\pm$0.3 & 0.2$\pm$0.3 & 0.07 & 41.2 & 35 & 2.2 \\[1mm]
        0717.15 & 109.3923348 & 37.738083 & 5.470 & 0.1$\pm$0.1 & ---$^{b}$ & 0.16 & 41.0 & 17 & 37.9 \\[1mm]
        0717.17 & 109.3914431 & 37.767048 & 5.458 & 0.8$\pm$0.4 & 1.0$^{+2.6}_{-1.0}$ & 0.04 & 41.4 & 69 & 4.0 \\[1mm]
        0717.25 & 109.4077276 & 37.742741 & 6.374 & 1.2$\pm$0.1 & ---$^{b}$ & 0.17 & 42.4 & 145 & 2.7 \\[1mm]
        0717.53 & 109.4128542 & 37.733804 & 6.348 & 0.2$\pm$0.1 & 0.23$\pm$0.01 & 0.004 & 40.7 & 17 & 34.8 \\[1mm]
        0717.59 & 109.37792 & 37.742851 & 5.637 & 0.2$\pm$0.4 & 3.7$\pm$0.1 & 4.09 & 41.2 & 1 & 2.0 \\[1mm]
        0744.17 & 116.202128 & 39.44821 & 5.718 & 3.5$\pm$0.7 & ---$^{b}$  & 0.19 & 41.2 & 26 & 1.2 \\[1mm]
        0744.60 & 116.22109 & 39.441155 & 5.899 & 0.4$\pm$0.2 & 0.2$^{+0.3}_{-0.2}$ & 0.15 & 41.6 & 57 & 3.8 \\[1mm]
        1149.51 & 177.413006 & 22.418862 & 6.694 & 0.3$\pm$0.5 & 0.6$\pm$0.2 & 1.34 & 41.7 & 17 & 1.8 \\[1mm]
        1149.67 & 177.412021 & 22.415777 & 6.621 & 1.1$\pm$0.3 & 1.3$\pm$0.1 & 1.57 & 41.7 & 6 & 1.8 \\[1mm]
        1347.4 & 206.895685 & -11.754647 & 6.598 & 1.6$\pm$0.7 & 1.6$\pm$0.8 & 0.4 & 41.9 & 76 & 1.4 \\[1mm]
        1347.25 & 206.8706306 & -11.753105 & 6.254 & 0.2$\pm$0.5 & 0.1$\pm$0.1 & 0.005 & 39.5 & 25 & 177.1 \\[1mm]
        1347.28 & 206.8750764 & -11.7588426 & 6.544 & 0.2$\pm$0.4 & 0.16 $\pm$0.01 & 0.03& 40.9 & 58 & 15.6 \\[1mm]
        1347.29 & 206.8865703 & -11.7620709 & 5.194 & 0.1$\pm$0.9 & 0.12$\pm$0.02 & 0.002 & 39.4 & 33 & 143.9 \\[1mm]
        1347.36 & 206.903015 & -11.750369 & 6.471 & 0.3$\pm$0.1 & 0.2$^{+0.4}_{-0.2}$ & 0.09 & 41.8 & 112 & 4.1 \\[1mm]
        1347.39 & 206.8886406 & -11.754282 & 5.286 & 0.2$\pm$0.5 & 0.4$\pm$0.2 & 0.03 & 40.0 & 9 & 21.2 \\[1mm]
        1347.45 & 206.8816569 & -11.761483 & 5.194 & 0.0$\pm$1.4 & ---$^{b}$  & 0.01 & 39.4 & 17 & 77.7 \\[1mm]
        1347.47 & 206.900859 & -11.754209 & 6.771 & 0.4$\pm$0.2 & 0.01$\pm$0.01 & 0.17 & 41.7 & 55 & 5.2 \\[1mm]
        1423.13 & 215.928816 & 24.083906 & 5.699 & 2.4$\pm$0.4 & 1.6$\pm$0.6 & 0.67 & 41.2 & 13 & 1.7 \\[1mm]
        1423.16 & 215.928929 & 24.072848 & 7.101 & 0.1$\pm$0.1 & ---$^{b}$  & 0.21 & 42.6 & 189 & 1.4 \\[1mm]
        1423.17 & 215.972591 & 24.072661 & 6.450 & 1.6$\pm$0.2 & 1.6$\pm$0.5 & 0.13 & 42.0 & 125 & 1.4 \\[1mm]
        1423.26 & 215.935869 & 24.078415 & 6.226 & 0.4$\pm$0.1 & 0.96$\pm$0.03 & 0.32 & 41.6 & 59 & 2.1 \\[1mm]
        1423.37 & 215.93618 & 24.074682 & 4.483 & 1.9$\pm$0.2 & 1.2$\pm$0.5 & 0.29 & 40.0 & 4 & 2.1 \\[1mm]
        1423.38 & 215.93638 & 24.057093 & 5.389 & 1.4$\pm$0.1 & ---$^{b}$ & 0.45 & 41.9 & 42 & 1.9 \\[1mm]
	2129.5 & 322.36419 & -7.701938 & 6.458 & 0.3$\pm$0.3 & ---$^{b}$  & 0.10 & 41.7 & 116 & 2.7 \\[1mm]
        2129.22 & 322.350939 & -7.693331 & 6.846 & 0.1$\pm$0.2 & ---$^{b}$  & 0.21 & 42.1 & 100 & 3.8 \\[1mm]
        2129.28 & 322.336347 & -7.696575 & 5.628 & 0.3$\pm$0.3 & ---$^{b}$  & 0.38 & 41.9 & 37 & 1.9 \\[1mm]
        2129.31 & 322.353238 & -7.697444 & 6.846 & 0.2$\pm$0.1 & 0.2$\pm$0.2 & 0.10 & 41.6 & 58 & 5.4 \\[1mm]
        2129.36 & 322.353941 & -7.681644 & 6.846 & 0.6$\pm$0.2 & ---$^{b}$  & 0.11 & 42.1 & 82 & 1.2 \\[1mm]
        2214.1 & 333.7234552 & -14.014297 & 5.847 & 0.7$\pm$0.1 & 1.2$\pm$1.2 & 0.15 & 42.4 & 407 & 1.5 \\[1mm]
        \hline \\[-3mm]
        \end{tabular}

    \end{tabular}

    \end{center}
$^a$ Corrected for best-fit magnification
$^b$ No \emph{HST}/WFC3 $F160W$ imaging, insufficient S/N, or reduction artifact

\end{table*}

\section{Results}
\label{results}
\subsection{Size and Pervasiveness of Ly$\alpha$-UV Offsets}
\label{LyaUVoff}
The filled gray histogram in the left panel of Figure \ref{fig:offsetfinal} shows the combined distribution of $|\Delta_{\rm{Ly}\alpha-\rm{UV}}|$ for the lensed and 
unlensed sample. The median value of the offset for the full sample, 0.61 kpc, is given in the top right of Figure \ref{fig:offsetfinal}, along with the 
effective 1$\sigma$ normalized median absolute deviation (NMAD; \citealt{hoaglin83}) scatter on 
the offset (0.68 kpc). The typical uncertainty on $|\Delta_{\rm{Ly}\alpha-\rm{UV}}|$ for the entire sample is 0.30 kpc, where uncertainties are derived 
only from the measurement process and do not include uncertainties associated with lensing or DAR corrections. While two-thirds 
of the entire sample have offsets significant at the $\ge1\sigma$ level and $\sim$40\% at the $\ge3\sigma$ level, at a cursory level, these offsets appear 
to be generally small. However, $>$10\% of our sample appear with Ly$\alpha$-UV offsets that exceed 2 kpc and the largest offsets reach up to nearly 
4 kpc, offsets that exceed the half-width of our DEIMOS slits and would have a considerable detrimental effect on our ability to detect these galaxies
in Ly$\alpha$ if this offset had been perpendicular to the spatial axis of our slits (see \S\ref{slitloss}). Additionally, as mentioned earlier, due to projection
effects, such measurements represent \emph{lower limits} to the true physical separation between the peak of the Ly$\alpha$ and UV continuum emission. We
note here, however, that it is unlikely that our observations missed even larger offsets, at least those that are somewhat aligned with the major axis of our
slits, as most galaxies were observed with slits designed to have $\sim$10-20 kpc of space on either side of the galaxy even after accounting for lensing effects.  

From Spearman correlation coefficient tests, we find no evidence of significant correlation between $|\Delta_{\rm{Ly}\alpha-\rm{UV}}|$ and either the 
$EW_{Ly\alpha}$ or the Ly$\alpha$ line luminosity. The lack of both is perhaps surprising since larger offsets should generally depreciate the measured line 
luminosity, or, equivalently, the $EW$ at fixed UV luminosity, if they occur isotropically. However, it is possible that we are only able to detect offsets 
in those galaxies for which the offsets are primarily
parallel to the major axis of the observed slits. Additionally, we do not have the ability to control for the intrinsic line luminosity, which further 
confuses this comparison. As such, we do not discuss this lack of correlation further opting instead to quantify slit losses for a variety of different spatial offsets 
in \S\ref{slitloss}. 

Additionally, we see no evidence within our own sample of evolution in $|\Delta_{\rm{Ly}\alpha-\rm{UV}}|$ as a function of redshift. A Spearman test returns 
no significant rejection of the null hypothesis that the two variables are uncorrelated, nor does a K-S test when breaking the sample into a lower and higher redshift
sample. As can be seen in the right panel of Figure \ref{fig:offsetfinal}, both small and large offsets are seen at essentially all redshifts probed by the combined
sample. 

In \citet{austin19b}, based on a sample of $\sim$300 LAEs in the redshift range $3\leq z \leq 5.5$ taken from the VANDELS survey \citep{mclure18, LPentz18b}, 
it was suggested that both the apparent and physical offsets between UV continuum and Ly$\alpha$ emission may decrease with increasing redshift. However, 
limitations in the statistical significance of the result occurred at the high-redshift end of the VANDELS sample ($z\ge4.5$). Here, we compare our projected
offset distribution with the equivalent quantity measured in \citet{austin19b} (see Figure 2 of \citealt{austin19b}) using identical redshift binning. We find 
no significant evidence of a difference between the distribution of projected offsets in our sample at $z\sim6$ and the full sample of \citet{austin19b}. 
Additionally, we do not find any significant differences between the two samples if we match the samples by UV luminosity or combine the two samples and bin 
by redshift. 

\begin{figure*}
\plotonekindaspecial{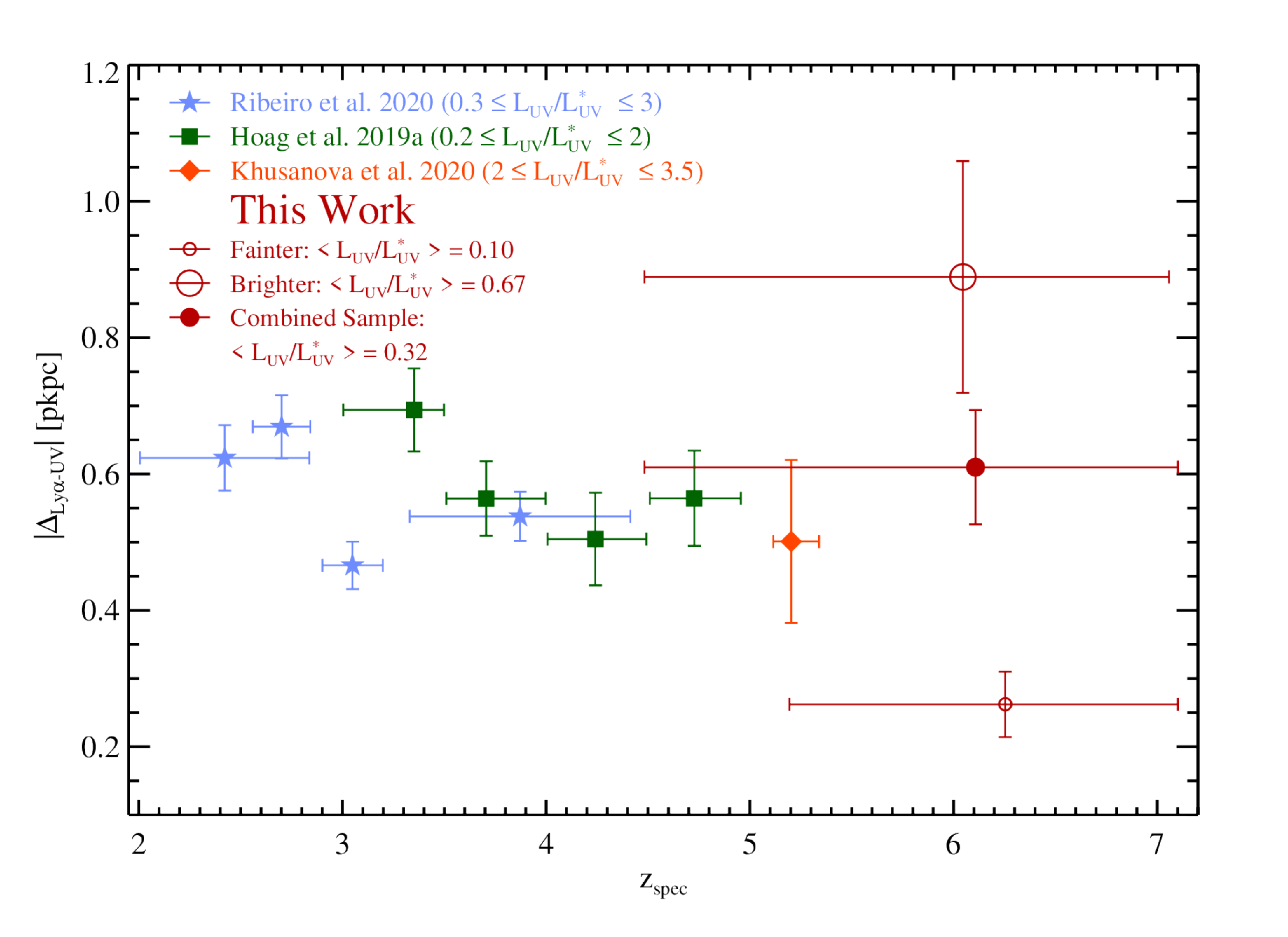}
\caption{Plot of the median \emph{projected one-dimensional} Ly$\alpha$-UV spatial offset as a function of median redshift for a compilation of all survey measurements from
$2 \le z \le 7$. The light blue stars show four equally-sized sub-samples at different redshift taken from \citet{bruno20}, the green squares show four sub-samples at different
redshift taken from the one-dimensional measurements from \citealt{austin19b}, and the orange diamond shows the median value from our own measurement of
six galaxies from the sample of \citealt{yana20}. The UV luminosity ranges of each sample are given in the legend at the top left. The red filled circle shows the median offset
measured in our sample, with the large and small open red circles indicated those values for our brighter ($\ge0.25 L_{UV}/L^{\ast}_{UV}$) and fainter ($<0.25 L_{UV}/L^{\ast}_{UV}$)
sub-samples. The median luminosities of our various (sub-)samples are given in the top left. In all cases, offset measurements represent lower limits to the true offsets in
these galaxies. Vertical error bars indicate the error on the median estimated by the NMAD/$\sqrt{n-1}$, where $n$ is the sample size (see \citealt{muller00, lem18}). The horizontal error bars indicate
the redshift extent of each (sub-)sample. No significant difference is found between the offset distributions or their medians for any of the samples at comparable
luminosities. A significant difference is observed, however, between the median offsets of our fainter and brighter samples.}
\label{fig:offsetvsz}
\end{figure*}

An identical comparison was also made to $\sim$900 LAEs observed at $2 \le z \le 6$ ($\tilde{z}=2.9$) as part of the VUDS survey\footnote{Note that this sample largely
does not include those galaxies presented in \citet{yana20} which, for completeness, were combined with our final sample and used as a high-redshift point of 
comparison for this exercise in addition to a comparison solely to our final sample.}. For this subset of VUDS 
LAEs, \citet{bruno20} present spatial offset measurements made on the observed continuum and the Ly$\alpha$ feature. 
In an internal comparison of these $\sim$900 LAEs, marginal evidence was found of redshift evolution in the projected one-dimensional spatial offsets, with 
increasing redshifts leading to slightly smaller offsets. Comparing the median projected offset found for the sample in \citet{bruno20} 
(0.60$\pm$0.05 kpc) to those found in our sample, 0.61$\pm$0.08 kpc, suggests no significant redshift evolution. This lack of significant difference persists when performing a K-S 
test on the two distributions, compare to sub-samples in different redshift bins, or (UV) luminosity match the data. We additionally compare both of these
samples and our own sample to the spatial offsets of the six galaxies at $5\le z\le 6$ from \citet{yana20} for which we could reliably measure an offset 
(see \S\ref{subsec:LPentz}). The median offset for this sample, 0.51$\pm$0.12 kpc, is statistically consistent with those of all other samples. Median spatial offsets 
of various redshift and luminosity sub-samples drawn from these four samples are plotted in Figure \ref{fig:offsetvsz}. With a large sample available at all redshifts 
from $2\le z \le 7$, the lack of significant difference in the offset distributions implies that whatever processes govern the offset of Ly$\alpha$ emission from the 
UV continuum are, at least, not a strong function of redshift.  

It does appear, however, within our own sample that the processes governing Ly$\alpha$ offsets are related to the UV luminosity of the host galaxy. 
In the right panel of Figure \ref{fig:offsetfinal} offsets are plotted as a function of $L_{UV}/L^{\ast}_{UV}$ and in the left panel the offset distribution is 
shown for galaxies that are brighter and fainter than $0.25L^{\ast}_{UV}$, which is roughly $M_{UV}\sim-19.5$. This splitting resulted in 36 galaxies in the 
brighter sample and 29 galaxies in the fainter sample. Note that the choice of this luminosity was made to create roughly equal-sized samples and 
none of our results change meaningfully if we delineate at a UV luminosity that is up to $\sim$50\% brighter or fainter than the fiducial luminosity.
Offsets generally appear to increase with increasing (UV) brightness, with a Spearman test returning evidence for a positive correlation ($\rho_{Spearman}=0.21$) at a $\sim$2$\sigma$ level
and a K-S test returning a rejection of the null hypothesis that the two distributions are drawn from a common underlying sample at the $\sim$3$\sigma$ level. 
Note that these statements are not particularly sensitive to the exact ranges chosen to define the two samples, as the Spearman test is independent
of this choice, and a K-S test rejects the null hypothesis at the $\ga$2$\sigma$ level irrespective of the exact choice of $L_{UV}/L^{\ast}_{UV}$ used to delineate between the 
faint and bright samples. While offsets for galaxies of all UV luminosities generally tend to be small, the vast majority ($\sim80$\%) of the larger offsets ($\ge1.5$ kpc) 
are observed in the brighter sample. Further, the median offset for the brighter sample, 0.89$\pm$0.18 kpc, is approximately three times larger than that of the fainter sample, 
0.27$\pm$0.05 kpc. Though the dependency of $|\Delta_{\rm{Ly}\alpha-\rm{UV}}|$ on UV luminosity varies with the comparison being made and the exact 
sample definition, the positive correlation between the two ranges from marginally to highly significant across all combinations and always favors larger offsets 
in UV-brighter galaxies. This relationship also appears to persist at approximately the same level of significance when estimating the instrinic offset distributions 
for the two luminosity sub-samples (see \S\ref{intrinsic} and Appendix \ref{appendix:B}).

These results remain broadly unchanged if we excise the most highly-magnified galaxies from our sample ($\mu>50$). These are the galaxies in our sample where 
the lensing approximation given in \S\ref{sec:UVcent1} is likely to have the largest effect. However, even after excising the highly-magnified galaxies, it 
is still possible that the difference we observe here could be attributed, at least in part, to small-scale lensing effects rather than any true physical differences between 
the samples. This would be the case only if this lensing approximation is not valid \emph{on average} for the $\sim25$ remaining magnified galaxies. 
Unfortunately, the equivalent exercise of comparing offsets to comparably UV-fainter 
galaxies as our lower-luminosity sample cannot be performed on the data in \citet{austin19b} 
or \citet{bruno20}, as both studies only probe galaxies with UV luminosities that are comparable to our brighter sample. Larger samples at lower-luminosity 
as well as better modeling of local lensing effects will be necessary to disambiguate the possible origins of this difference.

In a standard flux-limited survey in a blank field, in order to make a robust claim about differences observed as a function of galaxy brightness, 
it is necessary to mitigate issues related to Malmquist bias and differential selection effects relating to the difference in the intrinsic brightness of the
sources of interest. In a spectroscopic survey employing slit observations, it would additionally be necessary to discuss any differential loss of light between the two
populations as a resulting from the peculiarities of spectroscopic observations with finite-width slits. In the measurements presented here, this may be seen as particularly 
important, as it may be that we are simply missing the large-offset tail of 
lower-luminosity galaxies due to excess slit loss on a population that would, in a typical flux-limited field survey, be near the edge of detectability. 

However, one of the many virtues of a survey of lensed candidates is that the relationship between intrinsic brightness and the expected flux is modulated.
Additionally, we are binning galaxies by their intrinsic UV luminosity, whereas we are discussing the measurement of the Ly$\alpha$ line, a relationship 
which is further modulated by complicated processes internal to the emitting galaxies. In our own sample, while there is a strong and significant positive correlation between the magnification-corrected
Ly$\alpha$ line luminosity and intrinsic $L_{UV}/L^{\ast}_{UV}$ ($\rho_{Spearman}=0.59$, $p=2e-7$), there is, conversely, a strong and significant \emph{anti-}correlation 
between the observed Ly$\alpha$ line flux and intrinsic $L_{UV}/L^{\ast}_{UV}$ ($\rho_{Spearman}=-0.64$, $p=9e-9$). While large magnification values, values which 
would preferentially apply to the lower-luminosity sample, may result in pushing the Ly$\alpha$ emitting region out of the slit along the major axis in the case of
relatively short slits, the typical (median) slit length of our observations is 10.3$\arcsec$ with an NMAD scatter of 3.3$\arcsec$. For galaxies near the center of
the slit along the major axis, as, indeed, most targeted objects\footnote{Though this is not true for serendipitous detections, there are only three such
galaxies in the lower-luminosity sample ($\sim10$\% of the sample), and the few that are in our sample are also reasonably well situated in the center 
of the major axis of the slit \citep{spencer20}} are, 
as mentioned earlier, such a length allows for $\sim$10 kpc of space along the major axis at the median magnification of our lensed sample 
$\tilde{\mu}\sim3$. Thus, it is unlikely that differential observational issues are responsible for the lower $|\Delta_{\rm{Ly}\alpha-\rm{UV}}|$ values observed in the 
lower-luminosity sample. We also note here that slit position angles for the magnified sample were not purposely oriented to along the magnified major
axis of the galaxy, which would potentially differentially suppress or enhance our ability to detect offset Ly$\alpha$ for this sample relative to the unmagnified sample.


In \S\ref{restframeUVmorph} we measured the effective radii, $r_{e}$ for the 48 galaxies in our full sample with sufficiently deep \emph{HST}/$F160W$ imaging
to make a reliable measurement of this quantity. These 48 galaxies are comprised of 19 galaxies in the lower-luminosity sample and 29 in the higher-luminosity sample. 
Comparing $|\Delta_{\rm{Ly}\alpha-\rm{UV}}|$ to the measured $r_{e}$ values for the full sample, we find a reasonably strong and significant positive correlation 
($\rho_{Spearman}=0.35$, $p=0.014$) between the two quantities, which implies that larger offsets originate from galaxies with larger UV extents. This difference is also reflected in
the average (median) $r_e$ of the two luminosity sub-samples, $\widetilde{r_e}=0.32\pm0.06$ and $\widetilde{r_e}=0.95\pm0.11$ for the lower- and 
higher-luminosity sub-samples, respectively, a trend consistent with results on photometrically-selected candidates at similar redshift in the HFFs 
\citep{kawamata18}. Interestingly, these two values are consistent with the average $|\Delta_{\rm{Ly}\alpha-\rm{UV}}|$  observed for each 
sample and the ratio between the two is essentially identical to that of the $|\Delta_{\rm{Ly}\alpha-\rm{UV}}|$ values in the two samples. We will come back to this 
concordance in \S\ref{explaining}. 


\subsection{An Estimate of the One-Dimensional Intrinsic Ly$\alpha$-UV Offset Distribution}
\label{intrinsic}

In the previous section we discussed the distribution of projected one-dimensional offsets as measured. However, as discussed extensively in \citet{austin19b}, the measured
distribution differs from the intrinsic, though still one-dimensional and projected, distribution by the errors induced by the measurement process as well as any additional 
errors induced from the observations or analysis methods. In this section, we attempt to recover the intrinsic distribution of the offsets\footnote{From here on out we will 
refer to this distribution as ``intrinsic'' and ask the reader to keep in mind that it is still projected and one-dimensional and, thus, these intrinsic offsets 
are still lower limits to the true three-dimensional offsets.} by decoupling the spread induced through various errors from the intrinsic distribution. 

We attempt two methods to decouple the errors from the measured distribution. Both methodologies model the observed distribution as the convolution 
of two Gaussians one resulting from observational, analysis, and measurement error, given by: 

\begin{equation}
\Delta_{\rm{Ly}\alpha-\rm{UV}, \, err} = \frac{A^{\prime}}{\sqrt{2\pi}\sigma_{err}}e^{\frac{-(\Delta-\mu_{err})^2}{2\sigma_{err}^2}} 
\label{eqn:gauss1}
\end{equation}

\noindent where $A^{\prime}$ is the normalization, $\mu_{err}$ is the average offset, and $\sigma_{err}$ is the spread in the distribution induced by observational, measurement, 
and analysis errors. The second Gaussian results from the instrinsic Ly$\alpha$-UV offsets and is given by:

\begin{equation}
\Delta_{\rm{Ly}\alpha-\rm{UV}, \, int} = \frac{A^{\prime\prime}}{\sqrt{2\pi}\sigma_{int}}e^{\frac{-(\Delta-\mu_{int})^2}{2\sigma_{int}^2}} 
\label{eqn:gauss2}
\end{equation}

\noindent where $\mu_{int}$ and $\sigma_{int}$ are the average intrinsic offset and the intrinsic spread of the distribution, respectively. The measured distribution is then:

\begin{multline}
\Delta_{\rm{Ly}\alpha-\rm{UV}, \, err} \otimes \Delta_{\rm{Ly}\alpha-\rm{UV}, \, int} = \\ \frac{A}{\sqrt{2\pi}\sqrt{\sigma_{err}^2+\sigma_{int}^2}}e^{\frac{-(\Delta-\mu)^2}{2 (\sigma_{err}^2+\sigma_{int}^2)}}
\label{eqn:gauss3}
\end{multline}

\noindent where $\mu \equiv \mu_{err}+\mu_{int}$ (though see the note on $\mu_{int}$ later in this section). 

In this section we describe the first of the two methods, in which fits are performed to the observed distributions 
and the intrinsic distributions are inferred from those fits. A second method, in which we calculate the likelihood distribution of $\sigma_{int}$ is presented in Appendix \ref{appendix:B}. 
The method described in Appendix \ref{appendix:B} returns statistically consistent results to the method described in this section.

For this first method, rather than simply fit the expression in Equation \ref{eqn:gauss3} to the observed distribution of the various samples, 
we instead opted for a Monte Carlo process in which we tweaked each offset measurement by Gaussian sampling its formal random uncertainties 
1000 times and a fit is performed on each realization. For each fit, the larger of the two $\sigma$ values was adopted as that due to astrophysically-induced offsets 
(i.e., $\sigma_{int}$). This is likely a reasonable assumption, as the measured offsets are generally larger than our estimated errors. Additionally, as we will 
show later, the recovered $\sigma_{err}$ distribution is consistent with our average estimated error. 
Under the assumption that our errors are Gaussian and that we have accurately accounted for all sources 
of uncertainty, such a methodology allows us to create a probability density function (PDF) of the intrinsic distribution. From the resulting PDFs, we adopt the median
and the 16$^{\rm{th}}$/84$^{\rm{th}}$ percentiles as the values of $\sigma_{err}$ and $\sigma_{int}$ and their associated uncertainties, respectively. As an added benefit to 
this process, we can check whether there are hidden sources of observational or measurement error by examining the PDF of $\sigma_{err}$. If the median value of $\sigma_{err}$
differs appreciably from our median formal error (0.25-0.30 kpc depending on the sample) or its uncertainties are inconsistent with $\sigma_{err}$ being zero, as 
some Monte Carlo realizations are likely to stumble on the true distribution, then it is likely there are unaccounted sources of error. 

\begin{figure*}
\plotthree{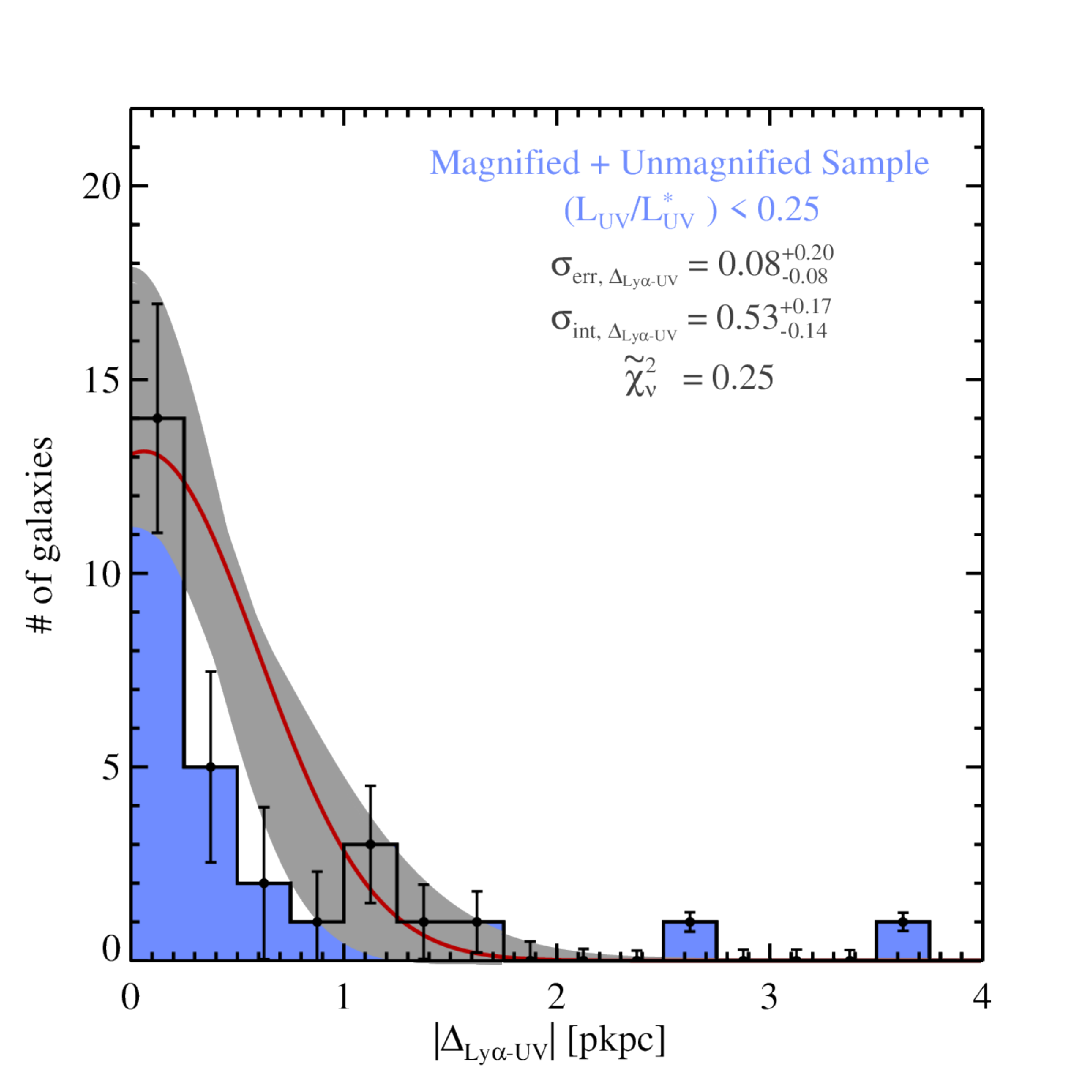}{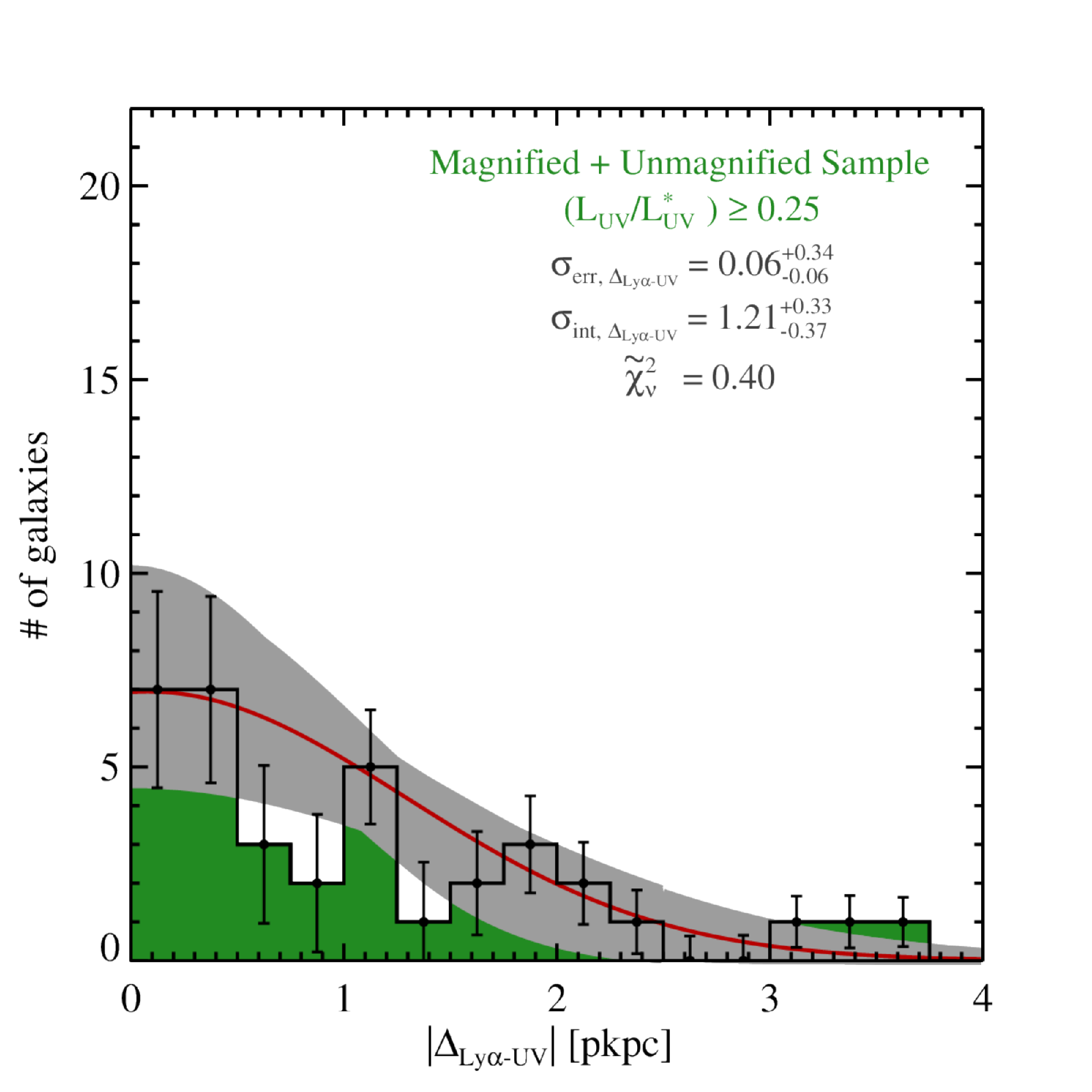}{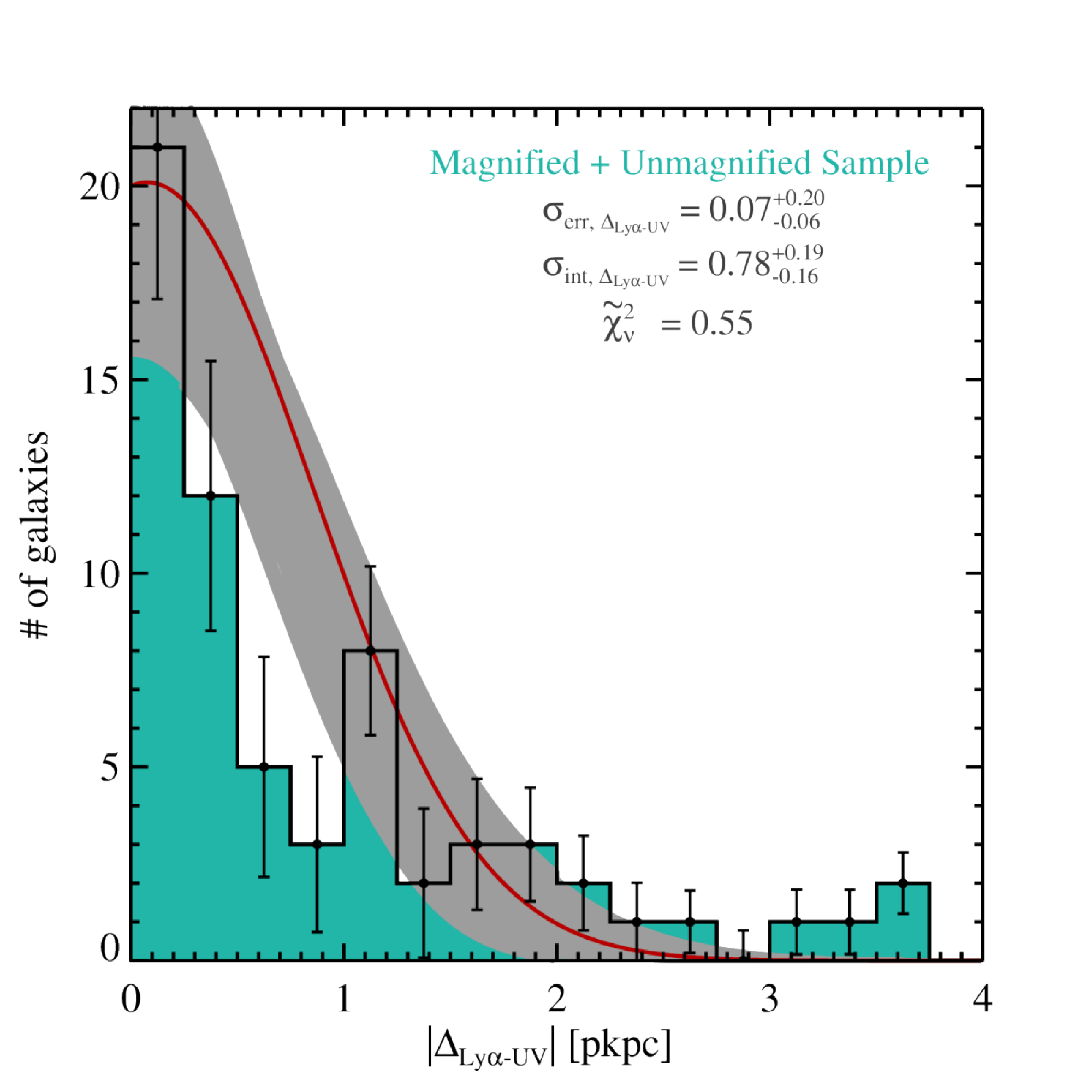}
\caption{\emph{Top Left:} Estimate of the instrinsic distribution of the one-dimensional projected Ly$\alpha$-UV offsets ($\Delta_{\rm{Ly}\alpha-\rm{UV}, \, int}$) for the low-luminosity
sample using the methodology described in \S\ref{intrinsic}. The solid histogram indicates the observed distribution of $\Delta_{\rm{Ly}\alpha-\rm{UV}}$, while the error bars show the effective
1$\sigma$ spread across the 1000 Monte Carlo realizations used to estimate the intrinsic distribution. The shaded region indicates the effective 1$\sigma$ spread of the convolved Gaussian best-fit
(see Equation \ref{eqn:gauss3}) to the observed distribution across all Monte Carlo realizations, while the solid red line indicates the median. The median values of the spread in the observed
distribution resulting from the error ($\sigma_{err}$) and from astrophysically-induced offsets ($\sigma_{int}$), along with their effective 1$\sigma$ bounds taken from the
16$^{\rm{th}}$/84$^{\rm{th}}$ percentile of all realizations are shown in the upper right. The median reduced $\chi^{2}$ of the best-fit models for all realizations is also shown.
\emph{Top Right:} Identical to the top left, but showing the results of this exercise for the higher-luminosity sample. \emph{Bottom:} Identical to the other two panels, but done
for the full sample.}
\label{fig:intrinsicoff}
\end{figure*}

In Figure \ref{fig:intrinsicoff} we plot the observed distribution of $|\Delta_{\rm{Ly}\alpha-\rm{UV}}|$ for the higher-luminosity, lower-luminosity, and full
with accompanying error bars reflecting the variation in the bin numbers across all 1000 Monte Carlo realizations. Overplotted on these observed distributions is the 
median fit to all Monte Carlo realizations along with a shaded region representing the $1\sigma$ region, i.e., the 16$^{\rm{th}}$/84$^{\rm{th}}$ percentiles of each
fit distribution. In each panel we indicate the median values of $\sigma_{err}$ and $\sigma_{int}$ and their associated uncertainties, as well as the median $\chi_{\nu}^2$ for 
the 1000 realizations. Note that, in each case, the mean of the convolved Gaussian given by $\mu$ in Equation \ref{eqn:gauss3} was statistically consistent with 
zero. Such a result is encouraging as, modulo statistical fluctuations in smaller samples, and under the assumption that
the model described by Equation \ref{eqn:gauss3} characterizes the distribution well, any deviations from zero should be entirely attributable to observational effects or 
measurement errors. I.e., in the large sample limit and adopting the assumptions of this exercise, $\mu$ should be equal to $\mu_{err}$ since 
the physical processes that are inducing the measured offsets, and thus dictate $\mu_{int}$, should have no preferred directionality in our slits. 
Also note, that despite the high degree of degeneracy between $\sigma_{err}$ and $\sigma_{int}$, the distributions of $\sigma_{err}$ and 
$\sigma_{int}$ across all Monte Carlo realizations do not appreciably overlap. 

It is clear from this exercise that the result found in the previous section regarding the difference between the offsets of the lower- and higher-luminosity
samples persists here. The two intrinsic spreads differ by $\sim$1.7$\sigma$ ($\sigma_{int}=0.53^{+0.17}_{-0.14}$ vs. $1.21^{+0.33}_{-0.37}$, respectively) and show
the same directionality as the measured difference. This is perhaps not surprising given the similar distribution of errors for both samples, but it serves to lend
credence to the inference made in the last section that offsets are generally larger in UV-brighter galaxies. Additionally, the 1$\sigma$ confidence interval for 
$\sigma_{err}$ for all three samples encompasses both our average formal uncertainty and zero, which suggests it is unlikely that there are additional hidden uncertainties
in our data. The full distribution, unsurprisingly, shows a intrinsic distribution intermediate to the lower- and higher-luminosity samples ($\sigma_{int}=0.78^{+0.19}_{-0.16}$).
Note that in all three cases with the possible exception of the higher-luminosity sub-sample, the Gaussian parameterization does poorly at properly accounting for the 
large offset tail of the distribution. 

In order to search for possible redshift evolution in the intrinsic distribution, we compare our recovered value of $\sigma_{int}$ for the higher-luminosity sub-sample
to a similar value computed by \citet{austin19b} for galaxies in five independent redshift bins running from $3 \le z < 5.5$. The higher-luminosity sub-sample is chosen
as a point of comparison because it contains galaxies with similar UV luminosities as those in \citet{austin19b}. While we do not adopt the complex 
forward-modeling approach of that work used to recover the intrinsic distribution, our recovered values still serves as a fair comparison under the assumption of 
azimuthal symmetry. Our estimated value of $\sigma_{int}$, though generally smaller than the equivalent values in \citet{austin19b}, differ from only their 
lowest-redshift measurement by $\ga3\sigma$. For the measurements of the remainder of the redshift bins the difference is $<1\sigma$. From this perspective,
there is also little evidence to suggest significant evolution in the intrinsic offset distribution as a function of redshift.

\subsection{Explaining the Ly$\alpha$-UV Offsets}
\label{explaining}
Several potential scenarios could be invoked to explain the behavior of increasing offsets in increasingly UV bright galaxies. The first, and perhaps most
obvious explanation, is that UV-brighter galaxies reside in more massive halos than their fainter counterparts (e.g., \citealt{mason15}), and, therefore 
are a more clustered population than their lower-luminosity counterparts (e.g., \citealt{ania18}). Under such a scenario, rather than 
witnessing true intragalaxy offsets, the offsets measured in our sample would predominantly result from emission originating from multiple galaxies in
close proximity with differing Ly$\alpha$/UV luminosity ratios. However, as discussed in \S\ref{restframeUVmorph}, this is unlikely to be the case as 
most of the galaxies in our sample are observed to be isolated systems, at least to the depth of our rest-frame UV imaging. For those 
systems where multiple components were observed, it appeared unlikely that the offset was due to the presence of multiple components (see, e.g., the left 
panel of Figure \ref{fig:offsetexample}). 

These considerations do not preclude late-stage merging activity as the genesis for at least some of the measured offsets, and future work searching for 
evidence of such activity in their rest-frame UV
images would be necessary to make definitive claims. However, such late-stage merging activity would have difficulty in 
inducing the larger offsets observed in our sample and cannot explain those galaxies where Ly$\alpha$ appears to originate well away from the UV bright
portions of the galaxy (see, e.g., both panels of Figure \ref{fig:offsetexample}). This logic also applies for scenarios where violent disk instabilities 
(VDI; \citealt{mo98,genzel08,dekel09,inoue16}) induce large star-forming clumps that could, while being sub-dominant in the rest-frame UV, act as sites of 
copious Ly$\alpha$ production with little to no obscuring dust. Observational evidence of such UV-faint, Ly$\alpha$-bright internal structure
has, through lensing, been observed at high redshifts (see, e.g., \citealt{eros17a, eros17b, eros19}). For more modest offsets, however, it seems possible, 
given that galaxy size appears to scale with the size of the offset, that structure internal to the galaxy, whether induced by VDIs or by merging activity, 
is the genesis for at least some of the offsets seen in this sample. Future analysis involving a more careful morphological analysis of the internal structure 
of these galaxies, as in, e.g., \citet{bruno17}, would be needed to make definitive claims.

A final possibility related to merging activity is the presence of  
UV-faint or UV-dark Ly$\alpha$-emitting galaxies (e.g., \citealt{bacon15, maseda18, mary20}) in a pre- or early-stage merging configuration. However, 
such systems are fairly rare, with number densities of $\sim2-10\times10^{-5}$ Mpc$^{-3}$ at $3 \la z \la 7$ from extremely deep MUSE observations, and 
have typical Ly$\alpha$ line luminosities approximately three times lower than that of our combined sample. As there is no correlation between intrinsic line 
luminosity and observed offset, i.e., offsets are not preferentially seen in galaxies with fainter $L_{Ly\alpha}$, it is unlikely that such galaxies
are primarily responsible for the observed offsets. 

A second set of scenarios involve Ly$\alpha$ emission from infalling or outflowing H {\small I} gas. A scenario involving inflowing gas  
follows that proposed by \citet{rauch11}, in which a $z=3.344$ galaxy is observed with a spectrum characteristic of a damped Ly$\alpha$ (DLA) system 
at a spatial location coincident with the broadband rest-frame UV emission and a spatially asymmetric Ly$\alpha$ emission that protrudes $\sim$20 kpc 
from the UV center, whose main component was observed to be offset at level comparable to the largest offsets observed in our sample. This galaxy, which has an 
\emph{HST}/ACS $F814W$ magnitude of $m_{F814W}=26.3$ \citep{coe06} translating to $L_{UV}\sim$0.3$L^{\ast}_{UV}$ assuming a
beta slope of $\beta=-2$ and $L^{\ast}_{UV}$ of \citet{anykey15}, has a UV brightness comparable to the median $L_{UV}$ of our sample. 
Under the scenario proposed in that work, the gaseous halo was supposed to be split open by beamed ionizing radiation or other anisotropic 
process, or, alternatively, a part of the galaxy containing young stars might have been exposed through tidal interaction with an unseen companion. 
With the gas no longer being capable of resonantly scattering Ly$\alpha$ photons due to its removal or ionization, Ly$\alpha$ photons were free to 
propagate and scatter off of infalling filaments. Additionally, ionizing radiation is similarly able to escape through these channels resulting in 
Ly$\alpha$ fluorescence off of infalling or nearby gas (see, e.g., \citealt{masribas16}). Such cavities are similarly suggested to induce
Ly$\alpha$ and Lyman continuum escape in the rare class of Lyman continuum emitting galaxies observed at high redshifts (see, e.g., 
\citealt{alice16,eros16, eros18,eros20b}). 

This phenomenon is potentially more likely to occur 
in the brighter sample due to the potential excess clustering, or, as we will discuss below, the potential for stronger stellar feedback. 
In addition, any activity due to an active galactic nuclei (AGN) could help facilitate and enhance this effect (e.g., \citealt{windhorst98}). 
While the Ly$\alpha$ emitting regions of the galaxies observed in our sample do not appear as large as those seen in \citet{rauch11}, this lack does not 
necessarily preclude such a possibility. Deeper Ly$\alpha$ spectroscopy to observe the full spatial extent of the Ly$\alpha$ emitting region along with 
an estimate of the systemic redshift and, possibly, the nuclear activity of each galaxy through, e.g., the detection of the CIII] $\lambda$1907,1909\AA\ 
doublet, HeII $\lambda$1640\AA\ emission (e.g., \citealt{paolo13}), or, constraints on the systemic redshift the [CII] $\lambda$158$\mu$m transition 
(e.g., \citealt{LPentz16, marusa17, paolo20}) would be necessary to test this scenario more fully. Deep IFU data in which the astrometry is sufficiently precise 
to allow for the measurement of small offsets (e.g., \citealt{bacon15}) could be used as a basis for such a test if accompanied by measures of systemic redshifts through deep NIR 
spectroscopy or ALMA observations. 


Alternatively, or in addition, anisotropic outflows due to stellar feedback could entrain Ly$\alpha$ photons, either through resonant scattering
or fluorescence, or otherwise facilitate their escape from the galaxies by decreasing the covering fraction of H {\small I} gas through the opening 
of channels in the gaseous halo. It has been shown at $z\sim1-2$ that, in lower-mass star-forming galaxies such as those studied here, stellar feedback is 
capable of altering the velocity structure of \emph{in situ} gas \citep{hirtenstein19, pellix20} especially for galaxies with the highest star-formation
rates ($\mathcal{SFR}$). Additionally, pervasive large outflow velocities have been observed in $\sim$$L^{\ast}$ star-forming galaxies at $2\la z \la 6$ devoid
of AGN activity (e.g., \citealt{steidel10, marghe17, sugahara19, ginolfi20}). Such outflows result in extended halos of chemically-enriched gas 
expelled from the galaxy (e.g., \citealt{seiji20}). 

In order to estimate the average level of stellar feedback in the galaxies in our two sub-samples, we began by taking the 
median rest-frame UV absolute magnitude of our fainter and brighter sample, $M_{UV}=-18.33$ and $-20.49$, respectively, and converted them to an average 
$\mathcal{SFR}$ for each sample using the relation of \citet{jaacks12}:

\begin{equation}
\log(\mathcal{SFR}) = -0.41M_{UV}-7.77  
\label{eqn:SFRTIR}
\end{equation}
 
\noindent with a scatter of $\sim0.2$ dex. This relation is derived from a suite of hydrodynamical simulations of galaxies at similar redshifts to those in our sample 
and is broadly consistent with 
estimates from a variety of observational works on high-redshift galaxies (e.g., \citealt{debarros14, dirtyaf20})\footnote{While the level of concordance of the normalization 
with observational works depends on the exact star-formation history assumed, the slope of the \citet{jaacks12} relation is consistent with these works, 
which, for this exercise, is the relevant concern.} and yields values that, for the purposes of this exercise, do not meaningfully depart from more traditional 
estimates \citep{kenn98}. This relation assumes a stellar extinction of $E(B-V)_{\ast}=0.1$ for all galaxies, a value consistent with estimates of the average		
dust extinction properties of high-redshift galaxies (e.g., \citealt{anykey12,kuang16, victoria20}). The resultant average $\mathcal{SFR}$s are 0.6$\pm$0.3 and 
4.3$\pm$1.9 $\mathcal{M}_{\odot}$ yr$^{-1}$ for the fainter and brighter samples, respectively. Assuming outflow rates of atomic gas induced by stellar feedback scale 
linearly with $\mathcal{SFR}$ with only a mild stellar mass dependence (e.g., \citealt{fluetsch19, ginolfi20}, though see also \citealt{lillaura19} for an alternative view), the 
outflow rate in the brighter sample is $\sim7.5\times$ 
larger than the fainter sample. Similarly, under the assumption that the outflow velocity of atomic gas scales as $\sqrt{\mathcal{SFR}}$ with, again, only a weak dependence on 
stellar mass (e.g., \citealt{heckman15,cicone16}), the outflow velocity induced by stellar feedback in the brighter sample is $\sim2.75\times$ that of the fainter sample. 
While it is possible that dust content may increase with increasing UV-brightness for high-redshift galaxies, the increase is likely mild 
(e.g., \citealt{anykey11, anykey14,smit12,javier16,yana20}), and would only serve to enhance this disparity. 
These more powerful outflows would have the potential to push more gas to larger galactocentric distances, allowing for the possibility of larger observed 
spatial offsets of fluorescing gas. In addition, if such outflows were anisotropic, the ionizing radiation would likely preferentially beam in the direction of the 
outflow, due to the lower covering fraction of neutral gas in that direction, which would result in a natural conduit through which to light the interface between
the galaxy and its circumgalatic medium. The various relations used for this scenario, $M_{UV}-\mathcal{SFR}$, $\mathcal{SFR}-\dot{M}_{out}$, or $\mathcal{SFR}-v_{out}$, have
been shown to not vary strongly as a function of redshift, satisfying the requirement that the evolution in the process or processes inducing the offsets must be mild or 
nonexistent over the redshift range $2 \le z \le 7$ (see \S\ref{LyaUVoff}). 


Such scenario is also broadly consistent with the results presented in \citet{bruno20}. In that study, the percentage of galaxies with considerable Ly$\alpha$-UV 
offsets, classified as ``Offset'' emitters in that work, generally increased with increasing UV brightness (see their Figure 4) and the specific star formation rate 
of that class was, on average, the highest amongst all of the galaxy classes studied (though their $\mathcal{SFR}$ was statistically indistinguishable from the other classes). 
Additionally, while the velocity offset between systemic and the interstellar medium (ISM) lines was similar in the Offset class relative to all other
LAEs analyzed in that work, the velocity offset between the systemic and the Ly$\alpha$ line was the largest in the Offset emitters, suggesting that 
the Ly$\alpha$ emission originated from backscattering off of outflowing gas or from an anisotropic gas distribution allowing for Ly$\alpha$ fluorescence 
from background gas in close proximity. Note again, however, while that study probed four magnitudes in UV brightness, the faintest galaxies in \citet{bruno20}
would still be considered part of our brighter sub-sample, precluding a more comprehensive comparison.

\begin{figure*}
\plottwo{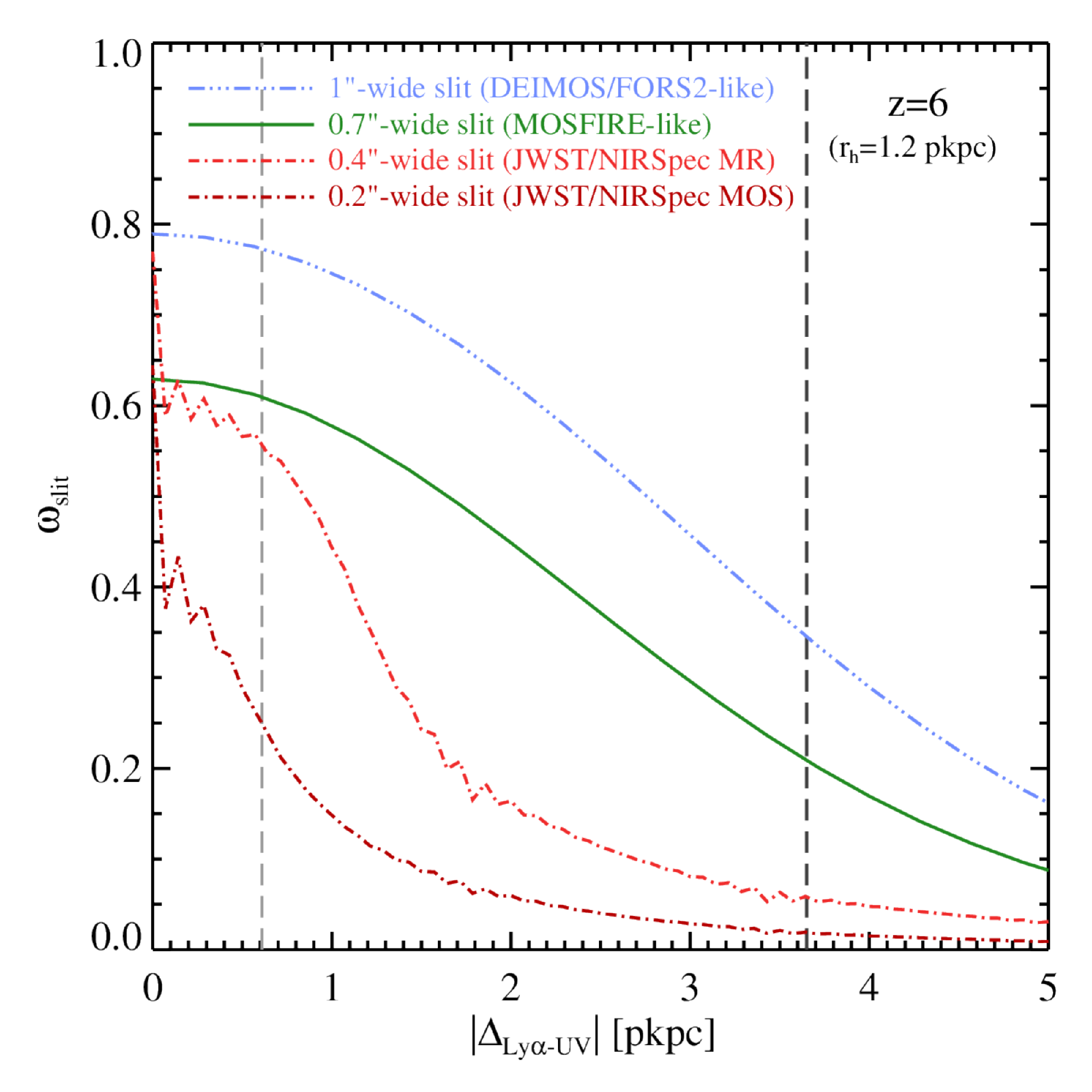}{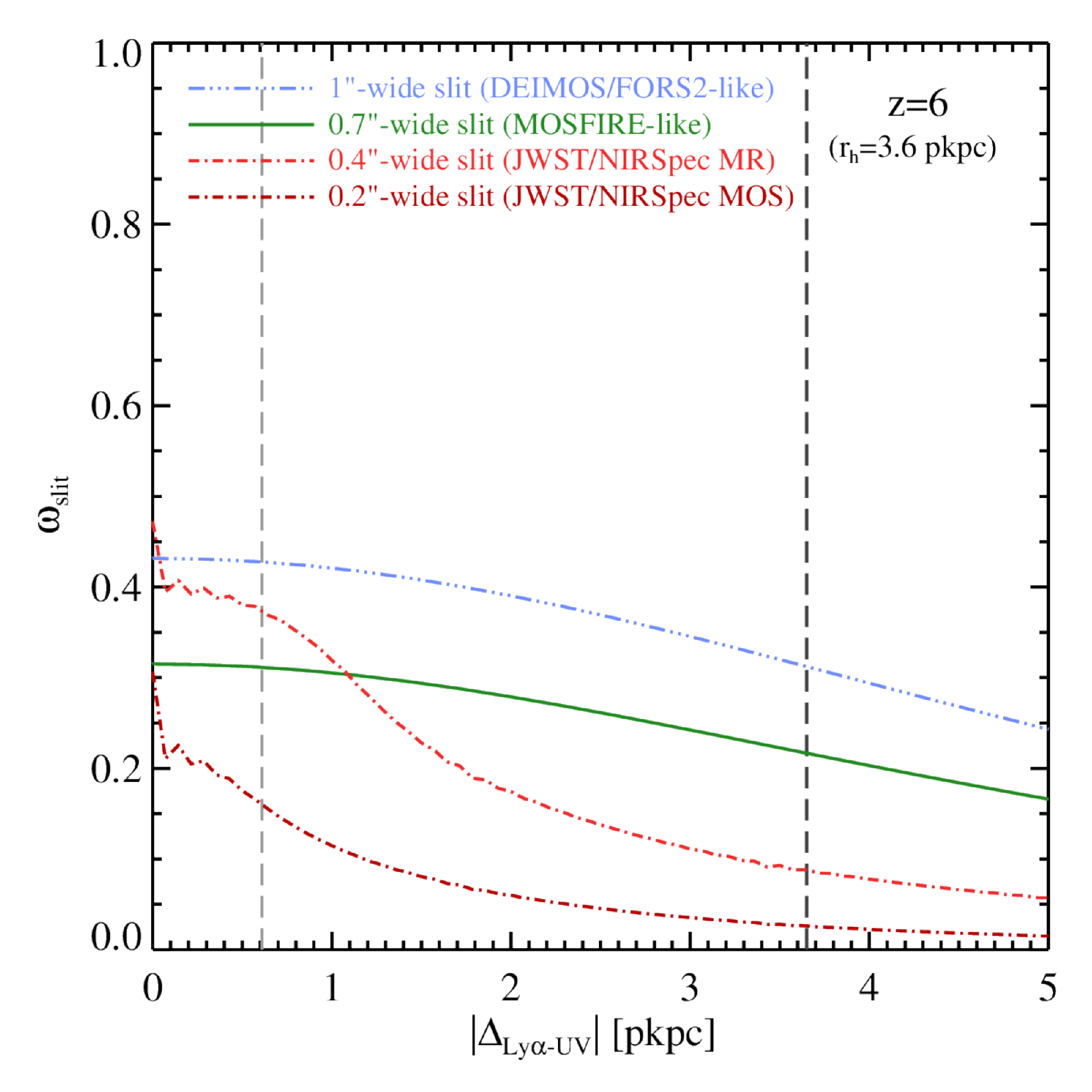}
\caption{\emph{Left:} Normalized slit throughput ($\omega_{slit}\equiv$1-slit loss) of Ly$\alpha$ emission of a simulated LAE as a function of the Ly$\alpha$-UV offset
($|\Delta_{\rm{Ly}\alpha-\rm{UV}}|$) for slit-spectroscopic observations using a variety of different instruments and telescopes. In each case, the slit
is assumed to be placed directly on the UV barycenter along both axes and the offset is assumed to be perpendicular to the major axis. The simulation 
uses the methodology described in \S\ref{slitloss} and \citet{lem09}. 
We adopt an exponential profile with $r_{h}=1.2$ kpc to characterize the Ly$\alpha$ emission and translate the angular offset to physical ones assuming
$z=6$. The light blue dot-dashed and solid green lines indicate $\omega_{slit}$ as a function of $|\Delta_{\rm{Ly}\alpha-\rm{UV}}|$ for ground-based slit 
observations employing 1.0$\arcsec$- and 0.7$\arcsec$-wide slits, respectively, under 0.8$\arcsec$ seeing. The orange and red dot-dashed lines show those
values for \emph{JWST}-NIRSpec medium resolution single slit and high-resolution multi-object spectroscopic observations, respectively. The light and dark
vertical dashed lines indicate the average and maximum $|\Delta_{\rm{Ly}\alpha-\rm{UV}}|$ for our sample. \emph{Right:} Indentical to the left panel, but the
emitting region of Ly$\alpha$ is now assumed to be three times the UV size of the galaxy.} 
\label{fig:slitloss}
\end{figure*}

\subsection{Induced Slit Losses}  
\label{slitloss}

While surveys of high-redshift LAEs are becoming common with grism observations (e.g., \citealt{pirzkal04, malholtra05, rhoads09, schmidt14, schmidt16, treu15, tilvi16, larson18}) 
and IFUs such as KCWI mounted on Keck {\small II} and MUSE mounted on the VLT Yepun Unit Telescope 
(e.g., \citealt{bacon15, bacon17, karman15, richard15, patoota17, patoota19, cai18, eros20b, nielsen20}), the targeting of high-redshift candidates with slit spectrographs remains a 
common observational strategy to census the high-redshift universe. Further, modern IFU studies from MUSE, for all but those with the deepest observations, lack the the astrometric 
precision needed to reliably measure $|\Delta_{\rm{Ly}\alpha-\rm{UV}}|$ at the level observed in the samples presented in this paper (see, e.g., \citealt{tanya19}). 
In a census of the high-redshift universe using slit spectroscopy, slits are almost always placed at the barycenter of the rest-frame continuum UV 
emission and integrated only long enough to detect galaxies if they have 
appreciable emission in Ly$\alpha$ (in the absence of other rest-frame UV emission lines). The resultant observations are used to derive the 
Ly$\alpha$ fraction at a fixed $EW$ limit for various sub-samples of galaxies or to otherwise probe the Ly$\alpha$ properties of the observed population to 
make inferences on the evolution of galaxy properties and the properties of the intergalactic medium at early times
(e.g., \citealt{ota08, LPentz11, stark11, treu13, tilvi14, debarros17, mason18, mason19, LPentz18, austin19a, spencer20}). As is 
discussed in detail in \citet{austin19b}, offsets between the UV and Ly$\alpha$ emission of target galaxies will, if severe, modulate the Ly$\alpha$ properties of 
the targeted population due to slit losses, which can, in turn, considerably modulate inferences on the nature of reionization. 

In order to quantify such losses for our sample, we relied on identical simulations to those described in \citet{lem09}. In these simulations, the Ly$\alpha$ halo is described 
by an exponential profile with $r_{h}=0.2\arcsec$, roughly the average (circularized) size of the UV continuum for galaxies at this redshift 
(e.g., \citealt{venemans05, overzier06,bruno16, LPentz18}). Note that none of the conclusions in this section change appreciably if we instead adopt the slightly 
smaller size characterized by the average measured $r_e$ value in our sample (see \S\ref{restframeUVmorph}). 
We will return to the case where the Ly$\alpha$ halo is considerably larger than the UV continuum later in the section. The simulated galaxies are well centered 
in a slit of length 10$\arcsec$, convolved with a Gaussian seeing kernel of varying sizes, and shifted along the minor axis in increments of 0.0125$\arcsec$. 
Slit widths are set to 1$\arcsec$, a typical width employed for FORS2/DEIMOS slits. The ratio 
between the light entering the slit and the total output of the galaxy, $\omega$, is considered the slit throughput, with 1-$\omega$ defining the slit loss. For a seeing 
kernel of full width at half maximum equal to 0.8$\arcsec$, a typical value for Mauna Kea in the $i$ band and also similar to the conditions required at Cerro Paranal for
our bright sample, the slit loss only increases\footnote{Here and throughout 
the section, unless noted otherwise, we are characterizing \emph{additional} slit losses induced by the spatial offset between the UV and Ly$\alpha$ emission. We do not additionally consider 
the slit losses that already occur when the Ly$\alpha$ emission is perfectly centered in the slit along its minor axis (see, e.g., \citealt{lem09} and Figure \ref{fig:slitloss}
for a characterization of absolute slit losses).} by $\sim$2\% for the average offset 
measured in our full sample, a negligible difference. This depreciation does not vary meaningfully if we use the average offset in the brighter or fainter samples observed 
in the previous section, nor if average projection effects are taken into account. It appears, as a whole, and differentially as a function of UV luminosity, the 
Ly$\alpha$ properties of our DEIMOS/FORS2 sample are largely unaffected by the average observed offsets. 


However, for individual offsets, the increase in slit loss, even with the relatively wide DEIMOS slits, can be devastating if the offsets unfortuitously manifest 
perpendicular to the major (spatial) axis of the slit. For the largest offsets observed in our combined sample ($\sim$3.5 kpc), offsets that are observed to varying 
degrees both in our fainter and brighter sample, additional slit losses exceed 50\% and increase to over 75\% if average projection effects are taken into
account. As such, it is clear that observations of large samples of galaxies at these redshifts are necessary to mitigate this stochasticity in order to draw  
robust conclusions about the nature of galaxy populations and reionization at these redshifts.

The above results remain broadly applicable for instruments that employ slightly smaller slit widths such as the Multi-Object Spectrograph for Infrared Exploration 
(MOSFIRE; \citealt{mclean12}). For typical MOSFIRE slit widths, 0.7$\arcsec$, the additional slit losses induced by the average offset in our sample are $\sim5$\%, again
a mostly negligible amount. However, large offsets, if perpendicular to the major axis of the slit, become an even larger hindrance, exceeding 70\% even under the assumption of
modest projection effects in our sample. These losses are further exacerbated in the case of targeted observations with the medium-resolution setting for the James 
Webb Space Telescope NIRSpec instrument, which employs 0.4$\arcsec$-wide slits. For the \emph{JWST}-NIRSpec simulations we do not convolve the simulated galaxy with 
an atmospheric seeing kernel, but rather were convolved with the NIRSpec $F110W$ PSF generated by the \textsc{Python}-based 
\textsc{WebbPSF}\footnote{\sloppy\url{https://www.stsci.edu/jwst/science-planning/proposal-planning-toolbox/psf-simulation-tool} tool. Additionally, we change the length
of the slit in the simulation to 3.65$\arcsec$, the nominal length for MR slit spectroscopy with NIRSpec\footnote{\url{https://jwst-docs.stsci.edu/near-infrared-spectrograph}},
though, in practice, this makes no appreciable difference.} In such observations, while additional slit losses are only $\sim$10-15\% for the average $|\Delta_{\rm{Ly}\alpha-\rm{UV}}|$ 
measured in our sample, the amount of Ly$\alpha$ entering the slit is severely reduced for the largest offsets observed in our sample, with $>90$\% additional slit 
losses when the offset is parallel to the slit minor axis and $\sim$95\% total slit loss. For large samples it will be important to account for this bias 
in order to properly quantify, e.g.,
Ly$\alpha$ escape fractions and to make inferences on kinematics from the Ly$\alpha$ line. 

The situation gets more dire for multi-object NIRSpec spectroscopy (NIRSpec-MOS),
which employs both narrower (0.2$\arcsec$) and shorter (1.38$\arcsec$) slits in its nominal mode. For such observations, even the relatively modest average 
$|\Delta_{\rm{Ly}\alpha-\rm{UV}}|$ measured in our sample results in additional slit losses of $\sim$60\% (75\% on an absolute scale) and slit losses approach complete 
at the largest offsets, with, \emph{at minimum} 98\% of the total Ly$\alpha$ flux lost for offsets perpendicular to the major axis of the slit. In the left 
panel of Figure \ref{fig:slitloss}, we plot the absolute slit losses as a function of $|\Delta_{\rm{Ly}\alpha-\rm{UV}}|$ for the two ground-based and two \emph{JWST} cases presented 
above.


If we instead assume in these simulations that the Ly$\alpha$ halo is considerably larger than the average size of the UV continuum (as in, e.g., 
\citealt{wisotzki16,leclercq17,erb18}), the additional
slit losses reported above are mitigated somewhat. For instance, if we assume that the half light radius of the Lya halo is three times that of the UV continuum, 
$r_{h,\,Ly\alpha}=0.6\arcsec$, the additional slit loss drops to 
negligible levels\footnote{Though, note that the larger halo increases \emph{total} amount of slit loss due to the larger size relative to the slit aperture.}.
However, in the case of the maximal offsets observed in our sample, \emph{even in the case of larger Ly$\alpha$ halos, the additional slit losses reach 60-70\% for
JWST-NIRSpec MR observations} if the offsets are parallel to the minor axis of the slit. Similarly, \emph{the slit losses remain near total for NIRSpec-MOS 
observations for galaxies with the largest offsets, with, at minimum, $\sim$97\% of the total Ly$\alpha$ flux incident lost, and 85\% of total 
for the average observed offset.} Clearly, regardless of the properties of the Ly$\alpha$ halos 
at high redshift, such offsets are crucial to quantify for observations of individual galaxies if the Ly$\alpha$ line is to be meaningfully used to place 
physical constraints.

\section{Conclusions}
\label{conclusions}
In this paper, we analyzed the projected spatial offsets of the UV continuum and the Ly$\alpha$ emission for a sample of 36 high-redshift ($5 \la z \la 7$) 
lensed LAEs taken from \citet{spencer20} and 30 high-redshift non-lensed LAEs drawn from \citet{LPentz11} and \citet{LPentz18}. This sample spans more than
three orders of magnitude in intrinsic UV brightness and represents the first such measurements performed on galaxies at these redshifts. The measured spatial offsets 
of these samples were compared internally for evidence of dependence on various physical parameters. A comparison was also made to a variety of 
lower-redshift samples compiled by \citet{bruno20}, \citet{austin19b}, and \citet{yana20} to investigate the possible redshift evolution of 
these offsets over a broad baseline ($2\la z \la 7$). Various scenarios were explored for the observed offsets and simulated spectroscopic observations 
employing slits of various sizes appropriate for ground-based and future space-based spectrographs were performed to estimate the loss of 
Ly$\alpha$ light induced by these offsets. We list our main conclusions below:

\begin{itemize}

\renewcommand{\labelitemi}{$\bullet$}

\item While $\sim$40\% of the LAEs in our sample have observed UV continuum to Ly$\alpha$ projected spatial offsets, $|\Delta_{\rm{Ly}\alpha-\rm{UV}}|$, 
significant at the $\ge$3$\sigma$, the median offset was found to be relatively modest, $|\widetilde{\Delta}_{\rm{Ly}\alpha-\rm{UV}}|=0.61\pm0.08$ kpc. 

\item A small fraction of our sample, $\sim$10\%, exhibited Ly$\alpha$ offsets of $\ge2$ kpc, and the largest Ly$\alpha$ offsets were observed nearly 
4 kpc from the UV barycenters. 

\item Within our own sample, and in comparisons to large samples of lower-redshift LAEs, no significant evolution was observed in $|\Delta_{\rm{Ly}\alpha-\rm{UV}}|$
as a function of redshift over the redshift range $2\la z \la 7$. 

\item Splitting our sample into UV-brighter ($\widetilde{L_{UV}}/L^{\ast}_{UV}=0.67$) and UV-fainter ($\widetilde{L_{UV}}/L^{\ast}_{UV}=0.10$) sub-samples, a significant difference was 
observed in $|\Delta_{\rm{Ly}\alpha-\rm{UV}}|$, with UV-brighter galaxies exhibiting offsets $\sim$3$\times$ larger than their UV-fainter counterparts 
(0.89$\pm$0.18 vs.\ 0.27$\pm$0.05 kpc, respectively). Despite the heterogeneous selection methods and the complications of lensing differentially affecting the 
primary constituents of the two sub-samples, we found this difference to persist at the $\ga2-3\sigma$ level through a variety of complementary approaches to measuring the 
disparity in the offsets. As such, we argued that the differences observed in $|\Delta_{\rm{Ly}\alpha-\rm{UV}}|$ between these samples represent a true 
difference rather than an artifact of the differing observing strategies, measurements, or lensing effects. A reasonably strong positive correlation
was also observed between $|\Delta_{\rm{Ly}\alpha-\rm{UV}}|$ and the intrinsic UV half-light radii of galaxies in our sample. 

\item A variety of scenarios were discussed to explain the larger observed offsets in the brighter sample. Merging was found to be an unlikely cause, though
it is possible or even likely that late-stage merging activity and violent disk instabilities could be responsible for some of the modest offsets 
observed in our sample. Scenarios involving Ly$\alpha$ fluorescence and/or backscattering appeared the most consistent with the observations, especially for those
galaxies where the $|\Delta_{\rm{Ly}\alpha-\rm{UV}}|$ exceeded their UV size.

\item From mock spectroscopic observations it was shown that the relatively small average $|\Delta_{\rm{Ly}\alpha-\rm{UV}}|$ offsets observed in our sample 
resulted in small additional slit losses ($\sim$2\%) for a DEIMOS-like setup with 1$\arcsec$-wide slits, losses which did not meaningfully change if we instead 
used those average offsets measured at other redshifts or for the brighter or fainter sub-samples. Employing smaller slit widths, e.g., a MOSFIRE-like 
0.7$\arcsec$-wide slit or a \emph{JWST}-NIRSpec-like single 0.4$\arcsec$-wide slit, only increased the average additional slit losses moderately ($\sim$5-15\%). 
However, for galaxies observed with a nominal \emph{JWST}-NIRSpec MOS setup, with slit sizes of 0.2$\arcsec\times$1.38$\arcsec$, even the relatively modest
average offsets have a large effect, with slit losses increasing $\sim$60\%. Further, galaxies with the largest observed offsets were found to have their 
Ly$\alpha$ line fluxes depreciated $\ga$70\% for 0.7-1$\arcsec$-wide slits in cases where the offset is perpendicular to the major axis of the slit. 
For 0.2$\arcsec$- and 0.4$\arcsec$-wide slits, such offsets would cause a $>$95\% loss of flux, which effectively renders Ly$\alpha$ 
unobservable for all but the brightest galaxies if mitigating measures are not taken. Both slitless and IFU observations will be crucial to 
contextualize these observations on an individual and statistical level.  

\end{itemize}
Given the relatively small average offsets observed over a large redshift baseline, it is unlikely that inferences on the
Epoch of Reionization using the LAE fraction or other observed properties of Ly$\alpha$ (e.g., \citealt{mason18, austin19a}) are altered meaningfully by these offsets
as long as the samples used are large enough. However, 
in smaller samples or studies of individual galaxies, it is clear that such offsets can have a severe deleterious effect on attempts at accurately measuring Ly$\alpha$ properties. While 
the ability to measure rest-frame optical emission lines in galaxies at the Epoch of Reionization will come with the launch of \emph{JWST}, the Ly$\alpha$ properties of such galaxies
will still be crucial to constrain considerable aspects of both the internal physics of galaxies and that of their surrounding medium. As such, care 
must be taken to quantify UV-Ly$\alpha$ offsets properly and account for them in either the observational strategy or mitigate the confusion they induce 
when making inferences on such populations. 

\section*{Acknowledgements}

{\footnotesize
This material is based upon work supported by the National Science Foundation under Grant No. AST-1815458 to M.B. and grant AST1810822 to T.T., and by NASA through grant
NNX14AN73H to M.B. and grant HST-GO-13459 (GLASS) to T.T.. C.M. acknowledges support provided by NASA through the NASA Hubble Fellowship grant HST-HF2-51413.001-A awarded 
by the Space Telescope Science Institute, which is operated by the Association of Universities for Research in Astronomy, Inc., for NASA, under contract NAS5-26555. 
This work is based on data obtained from ESO program 190.A-0685. TT acknowledge support by NASA through grant ``JWST-ERS-1324''.
We thank an anonymous referee for their careful reading of the manuscript and several helpful suggestions. 
BCL gratefully acknowledges Olga Cucciati and Ricardo Amor\'{i}n for careful reading of the manuscript and for numerous suggestions and Dan Coe for providing help
with calculating magnifications for different lensing models in the HFFs. 
BCL also gratefully acknowledges the support of Lori M. Lubin in helping to provide a foundation that allowed for the writing of this 
paper. This paper is dedicated to the life of Paul Thompson, a friend and a mentor from whom the first author learned much. 
We also acknowledge the ASTRODEEP and CLASH teams for their work on generating photometric catalogs for many of these clusters.
This study is based, in part, on data collected at the Subaru Telescope and obtained from the SMOKA, which is operated by the Astronomy Data Center, National Astronomical Observatory of
Japan. This work is based, in part, on observations made with the Spitzer Space Telescope, which is operated by the Jet Propulsion Laboratory, California
Institute of Technology under a contract with NASA. Some portion of the spectrographic data presented herein was based on observations obtained with the European 
Southern Observatory Very Large Telescope, Paranal, Chile. The remainder of the spectrographic
data presented herein were obtained at the W.M. Keck Observatory, which is operated as a scientific partnership among the California Institute of Technology, the University of
California, and the National Aeronautics and Space Administration. The Observatory was made possible by the generous financial support of the W.M. Keck
Foundation. We thank the indigenous Hawaiian community for allowing us to be guests on their sacred mountain, a privilege, without which, this
work would not have been possible. We are most fortunate to be able to conduct observations from this site.}

\vspace{0.2cm}
\noindent \textbf{DATA AVAILABILITY}

\vspace{0.2cm}

\noindent The data used in this study will be shared upon reasonable request to the corresponding author.




\bibliographystyle{mnras}
\bibliography{Lyaoffset}

\appendix

\section{Corrections for Differential Atmospheric Refraction}
\label{appendix:A}

In \S\ref{sec:Lyacent1} it was discussed that a correction to the observed Ly$\alpha$ position for our magnified sample was required to account for the effects of 
differential atmospheric refraction (DAR). We describe that DAR correction here as well as DAR and astrometric effects on the estimate of the UV location of galaxies
in our magnified sample. 


For each mask that contained a LAE candidate used in our final sample, we compiled all slits that contained filler targets, typically potential cluster members,
with apparent magnitudes\footnote{These slits were typically outside the \emph{HST} coverage. While for such slits the magnitudes were measured on a wide variety
of observations across the different fields, these magnitudes were typically measured on some form of $R$- or $I$-band images.} brighter than $m_{AB}<22.5$. For 
every such target on a given mask, the two dimensional spectrum was collapsed along the spectral axis in eight individual mostly airglow-free 40\AA-wide windows 
running from $\sim$6600\AA\ to $\sim$ 9700\AA. For a given target, the spatial centroid was measured using the Gaussian fit method described above in each
of the windows where the spectrum was defined, with typical centroid uncertainties being smaller than 1/10th of a pixel. This process was visually supervised and 
any windows with strong artifacts, astrophysical issues (e.g., an emission line appearing in a window), or other issues that prevented the spatial location from 
being measured robustly were excised. For all remaining windows, a first-order polynomial was fit to the object location vs. the central
wavelength of the window using least-squares. Then, for each mask, the first-order term (i.e., slope) was averaged for all bright objects on that mask in order to parameterize 
the movement due to DAR as a function of wavelength. Between one and 49 bright objects per mask, with an average of 12, were used in this process.  
Despite our observations being extremely red for optical data, this correction could be somewhat large relative to the effect we are attempting to measure, 
with typical slopes of $\sim3\times10^{-4}$ pix \AA$^{-1}$. Finally, the raw Ly$\alpha$ spatial location for a candidate on a given mask was corrected using 
the measured slope and the difference between the Ly$\alpha$ wavelength and that of the central wavelength. Note that the formal uncertainty in the
slope in all cases is dominated by other sources of error and is not included in our final error budget. 

A second DAR issue arises when attempting to estimate the UV location of the magnified sample, as none of the galaxies in the magnified sample had 
sufficiently strong continuum in the DEIMOS observations. As discussed in \S\ref{sec:UVcent1}, the DEIMOS design software, \textsc{dsim}, provides a predicted
location of an object at the central wavelength of the observations. However, the true location of an object can differ slightly due to observations being taken
under different conditions and at different times than those used to generate the design file.
In order to test the magnitude of this difference, we compared the predicted object location, called ``\textsc{CAT\_OBJP}'' in the header of the \textsc{spec2d} data
products, with the measured spatial centroid of each of the $\sim$400 bright objects as described in \S\ref{sec:Lyacent1}. This spatial centroid was estimated at 
the spectral window that most closely matches that of the central wavelength of the mask on which a bright object was observed. After accounting for an 
indexing issue that causes \textsc{CAT\_OBJP} to
overestimate the pixel location by $\sim$1 pixel, the median of the difference between \textsc{CAT\_OBJP} and the measured location was zero with a normalized median
absolute deviation (NMAD; \citealt{hoaglin83}) scatter of $\sim1.5$ pixels\footnote{While the scatter here could be thought of as a systematic uncertainty, none of the
main conclusions of this paper are meaningfully changed if we use a per-mask correction to the UV spatial location.}. The UV location for each candidate was 
then set to the index-corrected \textsc{CAT\_OBJP} value.

For the three candidates that were in our sample that were not targeted but which were fortuitously subtended by one
of our slits (``secondaries'' adopting the terminology of \citealt{spencer20}), the UV object location is not estimated by the mask design software.
In order to estimate a UV location that is methodologically consistent, we first estimated the location of the secondary from its
($\alpha$,$\delta$) in the ASTRODEEP catalog. This value was then corrected to an equivalent \textsc{CAT\_OBJP} by using the median of the offset between \textsc{CAT\_OBJP}
and the location estimate using the $\alpha$/$\delta$ of all targets on a given mask. This equivalent \textsc{CAT\_OBJP} was then treated the same as
the \textsc{CAT\_OBJP} values for the targeted candidates. An additional consideration in this process is the precision of the astrometry for all targets
and secondaries, as any issues with the astrometry will manifest themselves in the expected location of the candidate galaxy. However, comparing our original design
$\alpha$/$\delta$ values for all targeted bright objects ($m_{AB}<22$) across all fields to astrometric standards (e.g., GAIA2, SDSS, USNO), the NMAD scatter
in $\alpha$/$\delta$ is $\sim0.04\arcsec$, small enough to ignore for this analysis.

\section{Alternative Methodology to Infer the Intrinsic 1d Offset Distribution}
\label{appendix:B}

\begin{figure}
\plotone{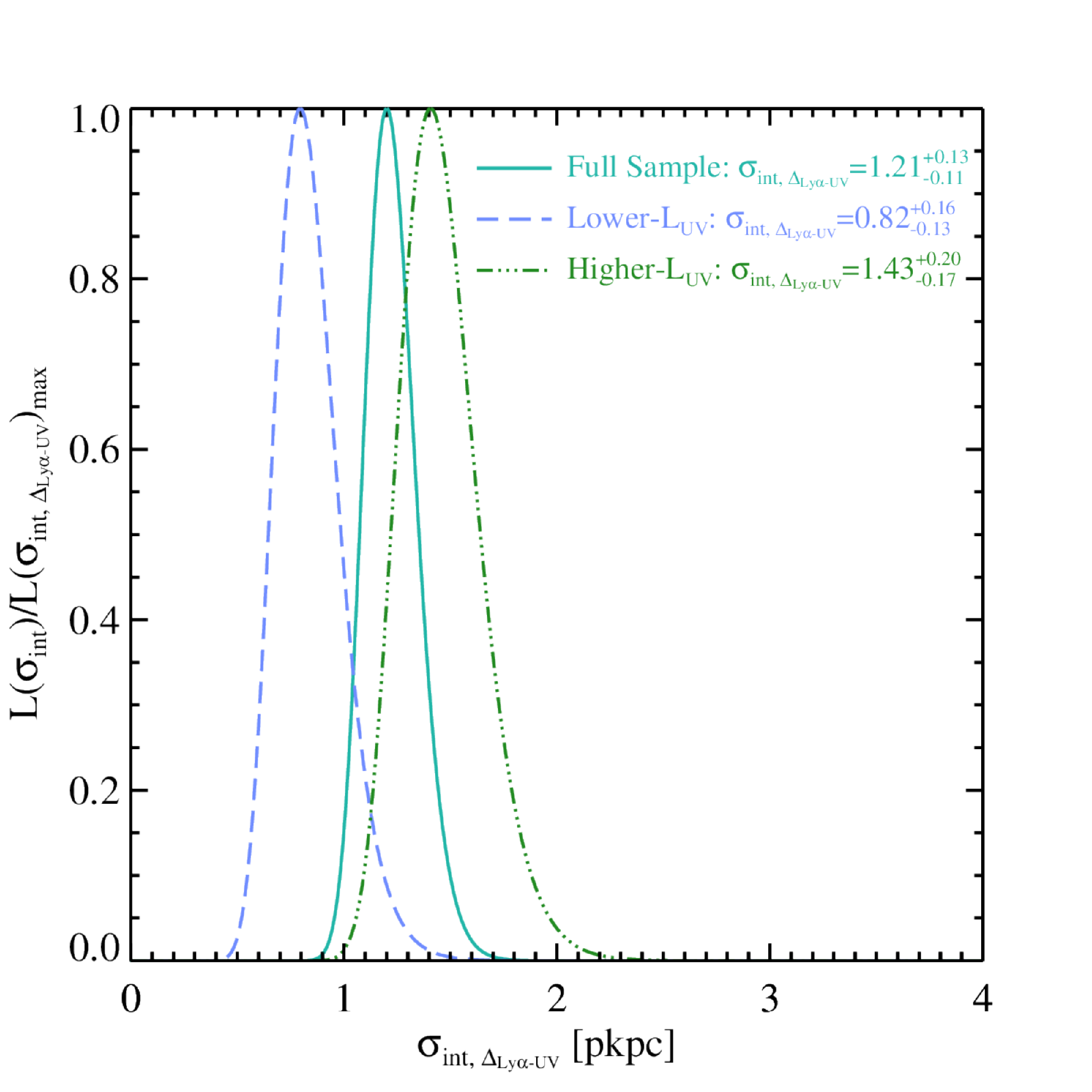}
\caption{Likelihood distribution of the intrinsic projected one-dimensional spread of $\Delta_{\rm{Ly}\alpha-\rm{UV}}$ for the full (solid teal), lower-luminosity (dashed light blue),
and higher-luminosity (dot-dashed green) samples. The the likelihood is calculated by the expression given in Equation \ref{eqn:gauss4}. The median of the likelihood distribution for each sample 
along with their corresponding effective $1\sigma$ confidence interval is shown in the top right. As in other analyses, offsets are found to be significantly higher for galaxies that are 
brighter in the UV.} 
\label{fig:likelihooddist}
\end{figure}

In this Appendix we adopt an alternative approach to inferring the intrinsic distribution of $\Delta_{\rm{Ly}\alpha-\rm{UV}}$ to that used in 
\S\ref{intrinsic}. As in \S\ref{intrinsic}, while we will refer to this distribution as intrinsic, we remind the reader that the values are still 
one-dimensional and projected, and thus represent a lower limit to the true three-dimensional offsets.  For this calculation we again assume that the 
underlying distribution can be characterized by a convolution of two Gaussians, one with spread $\sigma_{err}$ resulting from errors in the observation, 
measurement, and analysis processes, and one with spread $\sigma_{int}$ resulting from astrophysically-induced offsets. In this approach, we estimate the 
likelihood of given $\sigma_{int}$ for all measurements of a given sample by:

\begin{multline}
\Lagr(\sigma_{int}) = \sum\limits_{j=1}^N\frac{1}{\sqrt{2\pi}\sqrt{\sigma_{\Delta_{\rm{Ly}\alpha-\rm{UV}, j}}^2 + \sigma_{int}^2}}\\ e^{\frac{-(\Delta_{\rm{Ly}\alpha-\rm{UV}})^2}{ 2 (\sigma_{\Delta_{\rm{Ly}\alpha-\rm{UV}}, j}^2 + \sigma_{int}^2)}} 
\label{eqn:gauss4}
\end{multline}

\noindent where $\Delta_{\rm{Ly}\alpha-\rm{UV}, j}$ is the $j^{\rm{th}}$ measurement of the Ly$\alpha$-UV offset and $\sigma_{\Delta_{\rm{Ly}\alpha-\rm{UV}}, j}$ is its associated 
uncertainty. Note that unlike the approach in \S\ref{intrinsic}, we do not fit for $\sigma_{err}$, but rather use the estimated uncertainties of each 
$\Delta_{\rm{Ly}\alpha-\rm{UV}}$ measurement.
This likelihood is calculated for a finely-gridded array of $\sigma_{i}$ values running from $0 \le \Delta_{\rm{Ly}\alpha-\rm{UV}} \le 4$ for each sample. The resultant 
likelihood distributions on $\sigma_{i}$ are shown in Figure \ref{fig:likelihooddist} for the full, lower-luminosity, and higher-luminosity samples along with the median values 
and the 16$^{\rm{th}}$/84$^{\rm{th}}$ percentiles representing the effective 1$\sigma$ bounds of the distribution. The resultant values, while systematically larger than 
those in \S\ref{intrinsic}, are statistically consistent with the values measured with the method in that section for each sample. As in other analyses, we find that the 
most likely $\sigma_{int}$ value using this method for the lower-luminosity sample ($\sigma_{int}=0.82^{+0.16}_{-0.13}$) is lower than that of their higher-luminosity counterparts 
($\sigma_{int}=1.43^{+0.20}_{-0.17}$) formally at the $\sim$2.5$\sigma$ level, with very little overlap seen in the two likelihood distributions. The value for the higher-luminosity 
sample is now well in line with those measured by \citet{austin19b} at all redshifts save their lowest redshift measurement, again suggesting little to no redshift evolution of
$\Delta_{\rm{Ly}\alpha-\rm{UV}}$ for such galaxies. 

\bigskip
\bigskip
\bigskip

\noindent
\textbf{\large{Affiliations:}\\}
$^{1}$Department of Physics and Astronomy, University of California, Davis, One Shields Ave, Davis, CA 95616, USA\\
$^2$INAF--Osservatorio Astronomico di Roma, via di Frascati 33, I-00040, Monte Porzio Catone, Italy\\
$^3$Department of Physics and Astronomy, UCLA, Los Angeles, CA, 90095-1547, USA\\
$^4$W.\ M.\ Keck Observatory, 65-1120 Mamalahoa Hwy, Kamuela, HI 96743, USA\\
$^5$Harvard-Smithsonian Center for Astrophysics, 60 Garden St, Cambridge, MA, 02138, USA\\
$^6$Hubble Fellow\\
$^7$UCO/Lick Observatory, Department of Astronomy \& Astrophysics, UCSC, 1156 High Street, Santa Cruz, CA, 95064, USA\\
$^8$Leiden Observatory, Leiden University, PO Box 9513, 2300 RA Leiden, The Netherlands\\
$^9$Space Telescope Science Institute, 3700 San Martin Drive, Baltimore, MD 21218, USA\\
$^{10}$Leibniz-Institut f\"{u}r Astrophysik Potsdam (AIP), An der Sternwarte 16, 14482 Potsdam, Germany\\
$^{11}$INAF - Osservatorio di Astrofisica e Scienza dello Spazio di Bologna, via Gobetti 93/3 - 40129 Bologna - Italy\\
$^{12}$Max-Planck-Institut f\"{u}r Astronomie, K\"{o}nigstuhl 17, 69117 Heidelberg, Gemany\\
$^{13}$Aix Marseille Universit\'e, CNRS, LAM (Laboratoire d'Astrophysique de Marseille) UMR 7326, 13388, Marseille, France\\
$^{14}$N\'ucleo de Astronom\'ia, Facultad de Ingenier\'ia, Universidad Diego Portales, Av. Ej\'ercito 441, Santiago, Chile\\
$^{15}$Scientific Support Office, Directorate of Science and Robotic Exploration, European Space Research and Technology Centre (ESA/ESTEC), Keplerlaan 1, 2201 AZ Noordwijk, The Netherlands\\


\bsp    
\label{lastpage}
\end{document}